\documentclass[journal]{IEEEtran}

\usepackage{cite}

\ifCLASSINFOpdf
\else
\fi

\usepackage{array}
\usepackage[utf8]{inputenc}
\usepackage{amsmath,bm}
\usepackage{amsfonts}
\usepackage{amssymb}
\usepackage{amsthm}
\usepackage{graphicx,booktabs,longtable}
\usepackage{algorithm,algorithmic}
\usepackage{booktabs}
\usepackage[printonlyused, nolist]{acronym}
\usepackage{trfsigns}

\usepackage{tikz,pgfplots,pdflscape,pstricks}
\usetikzlibrary{arrows.meta,backgrounds,calc,decorations,patterns,positioning}
\usetikzlibrary{decorations.pathreplacing,angles,quotes,shapes,fit}
\tikzset{
	system/.style={draw,rectangle},%
	branch dot/.style={draw,fill,circle,inner sep=1pt},%
	binary op/.style={draw,circle,inner sep=0pt},%
	microphone/.style={draw,circle},%
}
\tikzset{
	punkt/.style={
		rectangle,
		rounded corners,
		draw=black, very thick,
		text width=6.5em,
		minimum height=2em,
		text centered},
	box/.style={
		rectangle,
		draw=black, thick,
		minimum height=2em,
		text centered},
	pil/.style={
		->,
		thick,
		shorten <=2pt,
		shorten >=2pt,}
}

\newlength\fwidth
\newcounter{MYtempeqncnt}

\newcommand{\transp}{^\text{T}}
\newcommand{\Freq}[1]{\underline{#1}}

\newcommand{\Timeix}{l}

\newcommand{\NumBlocks}{N}
\newcommand{\BlockIdx}{n}
\newcommand{\TriBlockIdx}{\BlockIdx}

\newcommand{\Freqix}{k}
\newcommand{\FreqixMax}{K}

\newcommand{\DFTLength}{K}

\newcommand{\W}{\mathbf{W}}
\newcommand{\WINST}{\breve{\Freq{\W}}}
\newcommand{\WIVA}{\Freq{\W}}
\newcommand{\Wcirc}{\mathbf{C}_{\W_{pq}}}

\newcommand{\y}{\mathbf{y}}
\newcommand{\x}{\mathbf{x}}

\newcommand{\Perm}{\mathbf{P}}
\newcommand{\yPerm}{\bar{\Freq{\y}}}
\newcommand{\xPerm}{\bar{\Freq{\x}}}
\newcommand{\WPerm}{\bar{\Freq{\W}}}

\newcommand{\Window}{\mathbf{U}}
\newcommand{\DFTWindowDR}{\Window_{D\times \DFTLength}^{1_{D}0}}
\newcommand{\DFTWindowRL}{\Window_{\DFTLength\times 2L}^{1_{2L} 0}}
\newcommand{\DFTWindowLD}{\Window_{2L\times D}^{1_{D} 0}}
\newcommand{\DFTWindowPLD}{\Window_{2PL\times PD}^{1_D0}}

\newcommand{\FourierMat}{\Freq{\mathbf{F}}_{\DFTLength}}
\newcommand{\FourierMatInv}{\FourierMat^{-1}}
\newcommand{\RootUnity}{\omega_{\DFTLength}}

\newcommand{\Walter}[1]{#1}
\newcommand{\RevTwo}[1]{#1}
\newcommand{\RevThree}[1]{#1}
\newcommand{\RevFour}[1]{#1}
\newcommand{\RevFive}[1]{#1}

\begin{document}
\title{A Unifying View on Blind Source Separation\\ of Convolutive Mixtures\\ based on Independent Component Analysis}

\author{Andreas Brendel ~\IEEEmembership{Student Member,~IEEE,} Thomas Haubner ~\IEEEmembership{Student Member,~IEEE,}\\ and Walter~Kellermann~\IEEEmembership{Fellow,~IEEE}% <-this % stops a space
\thanks{Andreas Brendel, Thomas Haubner and Walter Kellermann are with the chair of Multimedia Communications and Signal Processing, Friedrich-Alexander-Universit\"at Erlangen-N\"urnberg,
Cauerstr. 7, D-91058 Erlangen, Germany,
e-mail: \texttt{\{Andreas.Brendel, Thomas.Haubner, Walter.Kellermann\}@FAU.de}.}% <-this % stops a space
\thanks{This work was partially funded by the Deutsche Forschungsgemeinschaft (DFG, German Research Foundation) -- 282835863 -- within the Research Unit FOR2457 ``Acoustic Sensor Networks''.}% <-this % stops a space
}


\markboth{JOURNAL NAME}%
{Brendel, Haubner and Kellermann: A Unifying View on Blind Source Separation}



\begin{acronym}
	\acro{STFT}{Short-Time Fourier Transform}
	\acro{PSD}{Power Spectral Density}
	\acro{PDF}{Probability Density Function}
	\acro{RIR}{Room Impulse Response}
	\acro{FIR}{Finite Impulse Response}
	\acro{FFT}{Fast Fourier Transform}
	\acro{DFT}{Discrete Fourier Transform}
	\acro{ICA}{Independent Component Analysis}
	\acro{IVA}{Independent Vector Analysis}
	\acro{TRINICON}{TRIple-N Independent component analysis for CONvolutive mixtures}
	\acro{FD-ICA}{Frequency Domain ICA}
	\acro{BSS}{Blind Source Separation}
	\acro{NMF}{Nonnegative Matrix Factorization}
	\acro{MM}{Majorize-Minimize}
	\acro{SOS}{Second-Order Statistics}
	\acro{HOS}{Higher-Order Statistics}
	\acro{MIMO}{Multiple Input Multiple Output}
	\acro{SIR}{Signal-to-Interference Ratio}
	\acro{SDR}{Signal-to-Distortion Ratio}
	\acro{SAR}{Signal-to-Artefact Ratio}
	\acro{RTF}{Relative Transfer Function}
	\acro{ILRMA}{Independent Low-Rank Matrix Analysis}
\end{acronym}

\maketitle

\begin{abstract}
	\RevTwo{In many daily-life scenarios, acoustic sources recorded in an enclosure can only be observed with \RevFour{other} interfering sources. Hence}, convolutive \ac{BSS} is a central problem in audio signal processing. Methods based on \ac{ICA} are especially important in this field \RevTwo{as \RevThree{they require} only few and weak assumptions and \RevThree{allow} for blindness regarding the \RevThree{original} source signals and the acoustic propagation path.} \RevTwo{Most of the currently used algorithms belong to one of the following three families:} \ac{FD-ICA}, \ac{IVA}, and \ac{TRINICON}. \Walter{While} the relation between \ac{ICA}, \ac{FD-ICA} and \ac{IVA} becomes apparent due to their construction, the relation to \ac{TRINICON} is not \Walter{well} \RevTwo{established} yet. This paper fills this gap by providing an in-depth treatment of the common building blocks of these algorithms and their differences\Walter{, \RevTwo{and \RevThree{thus} provides a common framework} for all \RevThree{considered} algorithms}.
\end{abstract}

\begin{IEEEkeywords}
Blind Source Separation, Independent Component Analysis, Convolutive Mixtures, Indpendent Vector Analysis, TRINICON
\end{IEEEkeywords}

% \fi
\IEEEpeerreviewmaketitle

%%%%%%%%%%%%%%%%%%%%%%%%%%%%%%%%%%%%%%%%%%%%%%%%%%%%%%%%%%%%%%%%%%%%%%%%%%%%
\section{Introduction}
\label{sec:introduction}
%%%%%%%%%%%%%%%%%%%%%%%%%%%%%%%%%%%%%%%%%%%%%%%%%%%%%%%%%%%%%%%%%%%%%%%%%%%%
A real-world recording of an audio signal is \RevFour{often} a mixture of the desired source and undesired signals\RevFive{, e.g.,} in a cocktail party scenario. \RevThree{When} these recordings take place in acoustic enclosures, the observed microphone signals are affected by multiple reflections of the emitted \RevFour{signals}, i.e., reverberation. Hence, the observed mixture is \RevThree{typically} convolutive. Algorithms for convolutive \ac{BSS} \cite{makino_blind_2007} aim at \RevFour{decomposing} these mixtures into their \Walter{individual} \RevFour{source signals}. "Blind" refers to the fact that no or only little prior knowledge about the mixing process or the source signals is \Walter{exploited}.

There exists a rich literature \RevThree{on} convolutive \ac{BSS} algorithms in \RevThree{the} time domain as well as in \RevThree{the} frequency domain including different separation principles and cost functions (see \cite{pedersen_survey_2007} for an overview of convolutive \ac{BSS} methods). \RevTwo{Here, we will focus on multichannel algorithms, i.e., on approaches that exploit spatial diversity.}

%A large subset of \ac{BSS} methods are mask-based approaches \cite{deliang_wang_time--frequency_2008}, which exploit the W-disjoint orthogonality of speech signals in the \ac{STFT} domain \cite{rickard_sparse_2006}, e.g., DUET \cite{makino_duet_2007} or MESSL-type \cite{mandel_model-based_2010} algorithms. Recently, deep neural networks have been employed to learn a high-dimensional embedding of the speech signal mixture, which enables the estimation of a separating mask \cite{hershey_deep_2016}.
%
%Another type of learning-based approaches for source separation is \ac{NMF} \cite{lee_algorithms_2000,fevotte_nonnegative_2009}, which learns a dictionary representation of the source signals to be separated in a supervised or unsupervised manner. Several extensions of the single-channel \ac{NMF} approach to multichannel algorithms have been proposed \cite{ozerov_multichannel_2010,sawada_multichannel_2013}.
%
%Bayesian models and EM-based optimization strategies have been employed for convolutive source separation, which explicitly take prior knowledge about the \RevTwo{scenario} into account \cite{leglaive_multichannel_2016,kounades-bastian_variational_2016,brendel_spatially_2019}.
%
%Another option is to use the temporal structure or spatial distribution of the source signals, e.g., temporal changing source dominance \cite{laufer-goldshtein_source_2018} or a pretrained spatial model for the source positions \cite{taseska_spotforming:_2016}, to estimate relative transfer functions, which can be used to construct beamformers to extract or separate source signals.

A large class of convolutive \ac{BSS} algorithms use the statistical properties of the source signals for separation. \RevTwo{As the main assumption is the statistical independence of the source signals, such algorithms are usually called \acf{ICA}.} These statistical approaches can roughly \RevTwo{be} divided into two main subclasses, \ac{SOS}\RevThree{-} and \ac{HOS}-based approaches. \ac{SOS}-based approaches for separation of convolutive speech signal mixtures \RevThree{rely} on the nonwhiteness \cite{mei_blind_2004} or \RevThree{the} nonstationarity \cite{parra_convolutive_2000,weinstein_multi-channel_1993,aichner_time_2002} of speech \RevThree{or audio signals} or on both properties \RevFour{(see, e.g., \cite{tichavsky_fast_2011} for an overview)}. \ac{HOS}-based approaches use the assumption that the source signals are statistically independent, which can be \RevThree{expressed} by 4th-order moments \cite{comon_blind_2001,baumann_beamforming-based_2003}, nonlinear cross-moments \cite{charkani_convolutive_1999} or information theoretic measures, which are formulated w.r.t. the underlying source \acp{PDF}~\cite{bell_information-maximization_1995}.

In this contribution, we concentrate on convolutive \ac{BSS} methods based on \ac{ICA}, as a representative of the last mentioned group of algorithms. The first works on \ac{ICA} date back over thirty years, e.g.,  \cite{bell_information-maximization_1995}, and cover a broad variety of application scenarios including medical imaging, fMRI, EEG, telecommunications or stockmarket prediction \cite{hyvarinen_independent_2001}.

However, \ac{ICA} in its original form considers only instantaneous mixtures, which makes direct application \RevTwo{to} the separation of audio signals \RevThree{inappropriate} in most \RevThree{cases}\RevTwo{,} due to the \RevTwo{dispersive multipath propagation of sound waves in enclosures.} To address this issue, \ac{FD-ICA} has been proposed, which \RevTwo{models} convolutive mixing by estimating demixing filters \RevTwo{by instantaneous \ac{ICA} for each frequency bin in the \ac{STFT} domain \cite{smaragdis_blind_1998}.} However, \RevThree{then,} the permutation ambiguity of \ac{ICA}, i.e., \RevTwo{the uncertainty about the order of} the estimated signals at the outputs, appears in each frequency bin for \ac{FD-ICA}. \RevTwo{This yields the} well-known inner permutation problem, which has to be resolved to obtain \RevThree{acceptable} results \cite{sawada_robust_2004}.

To avoid the internal permutation problem, statistical coupling between the frequency bins \RevTwo{of the demixed signals} has been \RevTwo{enforced within} the class of \ac{IVA}-based algorithms \cite{kim_blind_2007}. Fast and stable update rules based on the \ac{MM} principle have been proposed (auxIVA) \cite{ono_stable_2011} and a unification of \ac{IVA} and \ac{NMF} has been \RevTwo{provided} \RevFour{as the \ac{ILRMA} method} \cite{kitamura_determined_2016}. \RevTwo{For solving} the external permutation problem, geometrically constrained \ac{IVA} algorithms have been proposed \cite{vincent_geometrically_2015, brendel_spatially_2019,brendel_journal} and a parametrization of the demixing matrices has been applied for signal extraction based on \ac{IVA} in \cite{koldovsky_gradient_2019}.

\ac{TRINICON}, a versatile framework for broadband adaptive \ac{MIMO} processing \RevTwo{modeling} the nongaussianity, nonwhiteness and nonstationarity of \RevTwo{signals like speech} has been proposed \RevTwo{in} \cite{buchner_generalization_2005,buchner_blind_2004} \RevTwo{and provides a unified framework for} \ac{SOS} and \ac{HOS}-based \ac{BSS} approaches. \ac{TRINICON} enables broadband acoustic \ac{BSS} \cite{aichner_acoustic_2007}, estimation of the direction of arrival \cite{lombard_tdoa_2011}, signal extraction \cite{reindl_minimum_2014} and joint separation and dereverberation \cite{buchner_trinicon_2010} \RevTwo{and, as a broadband technique \RevThree{based on a time-domain cost function}, inherently precludes the internal permutation problem.} Further, \RevTwo{the time-domain formulation ensures a} linear convolution model \RevThree{for the demixing - corresponding to the convolutive mixing of the acoustic sources -} instead of a circular convolution model as \RevTwo{intrinsic to} \ac{FD-ICA} and~\ac{IVA}.

This paper provides \RevTwo{a hitherto unavailable} \Walter{analytical, in-depth} comparison of the \RevTwo{three} most popular \ac{ICA}-based convolutive \ac{BSS} approaches \RevThree{for audio signals,} \RevTwo{\ac{FD-ICA},} \ac{IVA} and \ac{TRINICON}. \RevTwo{While} \cite{aichner_acoustic_2007} provides a comparison of \RevThree{\ac{IVA} and \ac{TRINICON} for} gradient-based updates, \RevTwo{a comparison for} the state-of-the-art optimization \RevTwo{methods} of \ac{IVA}, e.g., \ac{MM}-type update rules, and \RevTwo{various} versions of the gradient (holonomic or nonholonomic gradient) or Newton-type optimization strategies \RevTwo{are \RevThree{still} missing}. \RevTwo{Furthermore, a comparison of the update rules \RevFour{as in \cite{aichner_acoustic_2007}} does not explicitly show \RevThree{the underlying} model assumptions.} Therefore, a discussion of the differences of \RevTwo{these} approaches \RevFour{resulting from their different} cost functions, which \RevTwo{can be considered as} the most general approach for comparing \RevTwo{these} methods, is \RevTwo{provided in the sequel}. \RevTwo{We} will \RevTwo{establish} a common basis for the derivation of these algorithms and highlight corresponding steps in their derivation. \RevTwo{E}specially the in-depth discussion of the relation \RevTwo{of the \ac{IVA} cost function} to the \ac{TRINICON} cost function is deemed beneficial \RevTwo{as it shows} that the \ac{TRINICON} cost function contains the \ac{IVA} cost function as a special case. Furthermore, we will provide an analysis of the demixing models \RevTwo{underlying} the considered approaches. \RevTwo{Finally, \ac{IVA} and \ac{TRINICON} are compared by experiments using measured \acp{RIR}. Furthermore, \RevThree{benchmark results} using oracle \acp{RTF} representing a linear or a circular convolution are provided.}

The remainder of the paper is structured as follows: 
%Section~\ref{sec:notation} introduces notations used throughout the paper. 
The \RevTwo{assumed signal} model is \RevTwo{introduced} in Section~\ref{sec:mixing_model}. The cost functions of the three considered convolutive \ac{ICA}-based \ac{BSS} approaches \ac{FD-ICA}, \ac{IVA} and \ac{TRINICON} are \RevTwo{formulated in a unified manner} in \RevTwo{Sections}~\ref{sec:FD-ICA}, \ref{sec:IVA} and \ref{sec:TRINICON}, respectively. \RevTwo{Thereby}, we emphasize corresponding steps in the derivations of the cost \RevTwo{functions} to highlight similarities between the algorithms. In Section~\ref{sec:relation_between_BSS_algorithms} the \RevTwo{relations} between the algorithms \RevTwo{are} \RevThree{shown} analytically. The identified differences are discussed in Section~\ref{sec:pros_cons} \RevTwo{and experimental results quantifying the impact of the identified differences are presented.}
%Experiments showing the performance of the algorithms in realistic scenarios are presented in Section~\ref{sec:experiments}. 
The paper is concluded \RevTwo{in} Section~\ref{sec:conclusion}.

%%%%%%%%%%%%%%%%%%%%%%%%%%%%%%%%%%%%%%%%%%%%%%%%%%%%%%%%%%%%%%%%%%%%%%%%%%%%
\section{\RevTwo{Signal Model}}
\label{sec:mixing_model}
%%%%%%%%%%%%%%%%%%%%%%%%%%%%%%%%%%%%%%%%%%%%%%%%%%%%%%%%%%%%%%%%%%%%%%%%%%%%

%---------------------------------------------------------------------------
\subsection{Notation}
\label{sec:notation}
%---------------------------------------------------------------------------
We \RevTwo{introduce} some notations and list the most important symbols in Tab.~\ref{tab:notations} for later reference.
\begin{table}
	\centering
	\begin{tabular}{ll}
		\toprule
		$\mathbf{I}$, $\mathbf{0}$ &  identity and all-zero matrix\\
		$\Timeix$ &  time index\\
		%		$\TimeixMax$ &  number of time steps\\
		$\NumBlocks$ &  number of signal blocks\\
		$\BlockIdx$ &  time frame index\\
		$\Freqix$ &  frequency index\\
		$\FreqixMax$ &  number of frequency bins, DFT length\\
		$\x$ &  vector of microphone signals\\
		$\y$ &  vector of demixed signals\\
		$\W$ & convolutive demixing matrix in time domain\\
		$\WINST$ & instantaneous frequency bin-wise demixing matrix\\
		 &\quad \RevFour{in STFT domain}\\
		$\WIVA$ & \ac{IVA} demixing matrix \RevFour{in STFT domain}\\
		$\tilde{\W}$ & extended demixing matrix \RevFour{in time domain}\\
		$p(\cdot)$ & \RevFour{Probability Density Function (PDF)}\\
		$P$ & number of source/microphone signals and output channels\\
		$D,L$ & parameters describing the signal lengths\\
		$\Window$ & window matrix\\
		$\FourierMat$ & DFT matrix for DFT length $\DFTLength$\\
		$\Perm$ & permutation matrix\\
		$\Wcirc$ & circulant matrix corresponding to Toeplitz matrix $\W_{pq}$\\
		$J$ & cost function\\
		\bottomrule
	\end{tabular}\vspace{5pt}
	\caption{\RevTwo{Notations used}}
	\label{tab:notations}
\end{table}
Scalar variables are typeset as lower-case letters, constant scalars as upper-case letters, matrices as bold upper-case letters and vectors as bold lower-case letters. \RevTwo{Underlined quantities} $\Freq{(\cdot)}$ denote \RevTwo{\ac{STFT}-domain} variables. \RevTwo{Subscripts} of identity or all-zero matrices indicate their \RevThree{size}, i.e., $(\cdot)_A$ indicates an $A\times A$ matrix and $(\cdot)_{A\times B}$ indicates \RevTwo{a matrix of size} $A\times B$. \RevTwo{Superscripts} of window matrices indicate their structure, i.e., $\Window^{1_A 0}$ denotes an identity matrix \RevTwo{of size $A\times A$} stacked on top of an all-zero matrix. $(\cdot)\transp$ and $(\cdot)^\text{H}$ denote transpose and Hermitian, respectively. Permuted matrices and vectors are marked with an overbar $\bar{(\cdot)}$. The expectation operator is denoted as \RevThree{$\mathcal{E}[\cdot]$}, the Kullback-Leiber divergence as $\mathcal{KL}\{\cdot\Vert\cdot\}$ and the differential entropy by $\mathcal{H}_{(\cdot)}$.

%---------------------------------------------------------------------------
\subsection{\RevThree{Convolution Model}}
%---------------------------------------------------------------------------
In this contribution we consider a set of $P$ acoustic \Walter{point} sources in a reverberant enclosure which are observed by $P$ microphones, i.e., a determined scenario. \RevThree{Disregarding} additive microphone noise, the $p$th microphone signal $x_p(\Timeix)$ \RevTwo{sampled} at time instant $\Timeix$ can be described \RevTwo{in the discrete-time domain as a superposition of the convolutional products} of the $P$ acoustic sources with the corresponding \acp{RIR}\footnote{\RevTwo{Note that a reasonable approximation of the \acp{RIR} can be obtained by a finite length $N_\text{RIR}$.}}
\begin{equation}
	x_p(\Timeix) = \sum_{q=1}^{P}\sum_{\kappa = 0}^{N_\text{RIR}\RevFour{-1}}h_{qp}(\kappa)s_q(\Timeix-\kappa)\RevTwo{,}
	\label{eq:mixing_model}
\end{equation}
where $h_{qp}(\kappa)$, $\kappa= 0,\dots,N_\text{RIR}-1$ are the coefficients of the \ac{RIR} from source $q$ to microphone $p$ and $N_\text{RIR}$ is the number of \ac{FIR} filter coefficients\Walter{, which are assumed to be time-invariant for the remainder of this paper}.

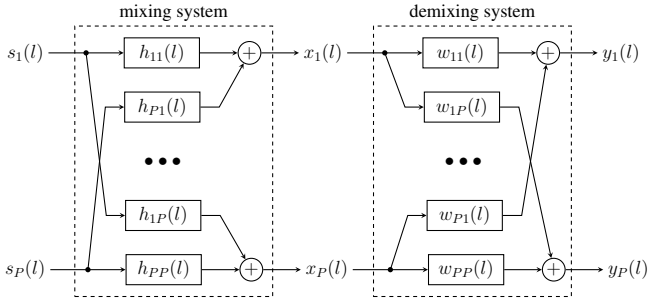
\begin{figure}
	\centering
	\scalebox{0.5}{
		%	\begin{tikzpicture}[thick]
%	
%	\node [] (s1) {$s_1(t)$};
%	\node [below= 1.5of s1] (sDots) {$\vdots$};
%	\node [below= 1.5of sDots] (sP) {$s_P(t)$};
%	
%	\node[right=of s1,branch dot] (dotS1) {};
%	\node[right=of sP,branch dot] (dotSP) {};
%	
%	\node[right=of dotS1,system] (h11) {$h_{11}$};
%	\node[below=-0.2 of h11] () {\scriptsize$\vdots$};
%	\node[below=0.5 of h11,system] (h12) {$h_{1P}$};
%	\node [below= 1.5of h11] () {$\vdots$};
%	
%	\node[right=of h11,binary op] (add1) {$+$};
%	\node [above=of add1] (noise1) {$n_1(t)$};
%	\node[right=of add1,branch dot] (x1) {};
%	\node[above=0.1of x1] () {$x_1(t)$};
%	\node [below= 1.5of x1] () {$\vdots$};
%	
%	\node[right=of dotSP,system] (h22) {$h_{PP}$};
%	\node[above=-0.1 of h22] () {\scriptsize$\vdots$};
%	\node[above=0.5 of h22,system] (h21) {$h_{P1}$};
%	
%	\node[right=of h22,binary op] (add2) {$+$};
%	\node [below=of add2] (noise2) {$n_P(t)$};
%	\node[right=of add2,branch dot] (x2) {};
%	\node[below=0.1of x2] () {$x_P(t)$};
%	
%	\draw (s1)--(dotS1);
%	\draw (dotS1)--(h11);
%	\draw (dotS1)|-(h12);
%	\draw (h11)--(add1);
%	\draw[out = 0,in = 90] (h12)-|(add2);
%	\draw[->] (noise1)--(add1);
%	\draw[->] (add1)--(x1);
%	
%	\draw (sP)--(dotSP);
%	\draw (dotSP)--(h22);
%	\draw (dotSP)|-(h21);
%	\draw (h22)--(add2);
%	\draw[out = 0,in = 90] (h21)-|(add1);
%	\draw[->] (noise2)--(add2);
%	\draw[->] (add2)--(x2);
%	
%	\end{tikzpicture}
\begin{tikzpicture}[thick,filter/.style={draw,minimum height=8mm,minimum width=20mm},%
binop/.style={draw,circle,inner sep=1pt},>=stealth, font = \Large]

% mixing system
\node[] (s1) {$s_1(\Timeix)$};
\node[below=5 of s1] (s2) {$s_P(\Timeix)$};

\node[coordinate,right=of s1] (branch-s1) {};
\node[coordinate,right=of s2] (branch-s2) {};
\fill (branch-s1) circle (2pt);
\fill (branch-s2) circle (2pt);

\node[filter,right=of branch-s1] (h11) {$h_{11}(\Timeix)$};
\node[filter,right=of branch-s2] (h22) {$h_{PP}(\Timeix)$};
\node[filter] (h21) at ($(h11)!0.75!(h22)$) {$h_{1P}(\Timeix)$}; % notation is consistent with previous literature, 
\node[filter] (h12) at ($(h11)!0.25!(h22)$) {$h_{P1}(\Timeix)$}; % but the internal notation with reversed indices is more intuitive for me
\node (dotsMixing) at ($(h11)!0.5!(h22)$) {$\bullet\bullet\bullet$}; % but the internal notation with reversed indices is more intuitive for me

\node[coordinate,left=0.5 of h21] (in-h21) {};
\node[coordinate,right=0.5 of h21] (out-h21) {};
\node[coordinate,left=0.5 of h12] (in-h12) {};
\node[coordinate,right=0.5 of h12] (out-h12) {};

\node[binop,right=of h11] (sum-s1) {$+$};
\node[binop,right=of h22] (sum-s2) {$+$};

\node[right=of sum-s1] (x1) {$x_1(\Timeix)$};
\node[right=of sum-s2] (x2) {$x_P(\Timeix)$};

% connections (mixing system)
\draw[->] (s1) -- (branch-s1) -- (h11);
\draw[->] (s2) -- (branch-s2) -- (h22);
\draw[->] (branch-s1) -- (in-h21) -- (h21);
\draw[->] (branch-s2) -- (in-h12) -- (h12);
\draw[->] (h11) -- (sum-s1);
\draw[->] (h22) -- (sum-s2);
\draw[->] (h21) -- (out-h21) -- (sum-s2);
\draw[->] (h12) -- (out-h12) -- (sum-s1);
\draw[->] (sum-s1) -- (x1);
\draw[->] (sum-s2) -- (x2);

% demixing system
\node[coordinate,right=of x1] (branch-x1) {};
\node[coordinate,right=of x2] (branch-x2) {};
\fill (branch-x1) circle (2pt);
\fill (branch-x2) circle (2pt);

\node[filter,right=of branch-x1] (w11) {$w_{11}(\Timeix)$};
\node[filter,right=of branch-x2] (w22) {$w_{PP}(\Timeix)$};
\node[filter] (w21) at ($(w11)!0.25!(w22)$) {$w_{1P}(\Timeix)$};
\node[filter] (w12) at ($(w11)!0.75!(w22)$) {$w_{P1}(\Timeix)$};
\node (dotsDemixing) at ($(w11)!0.5!(w22)$){$\bullet\bullet\bullet$};

%\node[left=of branch-x1] (s1) {$s_1$};
%\node[left=of branch-x2] (s2) {$s_2$};

\node[coordinate,left=0.5 of w21] (in-w21) {};
\node[coordinate,right=0.5 of w21] (out-w21) {};
\node[coordinate,left=0.5 of w12] (in-w12) {};
\node[coordinate,right=0.5 of w12] (out-w12) {};

\node[binop,right=of w11] (sum-x1) {$+$};
\node[binop,right=of w22] (sum-x2) {$+$};

\node[right=of sum-x1] (y1) {$y_1(\Timeix)$};
\node[right=of sum-x2] (y2) {$y_P(\Timeix)$};

% connections (demixing system)
\draw[->] (x1) -- (branch-x1) -- (w11);
\draw[->] (x2) -- (branch-x2) -- (w22);
\draw[->] (branch-x1) -- (in-w21) -- (w21);
\draw[->] (branch-x2) -- (in-w12) -- (w12);
\draw[->] (w11) -- (sum-x1);
\draw[->] (w22) -- (sum-x2);
\draw[->] (w21) -- (out-w21) -- (sum-x2);
\draw[->] (w12) -- (out-w12) -- (sum-x1);
\draw[->] (sum-x1) -- (y1);
\draw[->] (sum-x2) -- (y2);

\node[draw,inner sep=3mm,fit=(branch-s1) (h11) (h22) (sum-s1),dashed,label={[below,anchor=south] mixing system}] (mix) {};
\node[draw,inner sep=3mm,fit=(branch-x1) (w11) (w22) (sum-x1),dashed,label={[below,anchor=south] demixing system}] (demix) {};

\end{tikzpicture}}
	\caption{Exemplary \RevTwo{block diagram} showing the mixing and demixing system for a scenario with \RevTwo{$P$ sources and $P$} microphones.}
	\label{fig:blockdiagram_MixingDemixing}
\end{figure}

The set of FIR \RevTwo{models} \RevTwo{leading to} the microphone observations \Walter{is} called the mixing system. A broad class of \ac{BSS} algorithms aim at identifying a parametric demixing system, i.e., a set of \ac{FIR} filters which estimates \RevFour{--} \Walter{potentially} \RevTwo{reverberated} \RevFour{--} versions of the original speech signals from the observed mixture. The demixing system can \Walter{be} conveniently described by the so-called demixing matrix $\W$ \RevFour{which} is applied to the microphone signals \RevFour{to compute the demixed signals $y_q(\Timeix)$} \RevTwo{(see Fig.~\ref{fig:blockdiagram_MixingDemixing} for an illustration of the mixing and demixing \RevThree{systems})}. For generality \RevFour{and to cover both circular and linear convolution}, the application of the demixing matrix $\W$ is here abstracted by a function $\RevTwo{\text{conv}(\cdot)}$ so that we can write the separated output signals \RevFour{$y_q(\Timeix)$} as a function of the sensor signals $x_p(\Timeix)$\RevTwo{,} $\RevFour{p,q}\in\{1,\dots,P\}$\RevTwo{,} with filter length $L$
\begin{align}
	&[y_1(\Timeix),\dots,y_P(\Timeix)]\transp = \\
	&\qquad\qquad\qquad\qquad\RevTwo{\text{conv}}\bigg(\left \lbrace x_1(\Timeix'),\dots,x_P(\Timeix')\right\rbrace_{\Timeix+1-L\leq\Timeix'\leq\Timeix}\bigg).\notag
\end{align}
\Walter{As the mixing process is assumed to be time-invariant in this contribution, also the \RevTwo{optimum} demixing system \RevTwo{to be identified} can be assumed to be time-invariant here.}
%%%%%%%%%%%%%%%%%%%%%%%%%%%%%%%%%%%%%%%%%%%%%%%%%%%%%%%%%%%%%%%%%%%%%%%%%%%%
\section{\RevTwo{Frequency-Domain} Independent Component Analysis}
\label{sec:FD-ICA}
%%%%%%%%%%%%%%%%%%%%%%%%%%%%%%%%%%%%%%%%%%%%%%%%%%%%%%%%%%%%%%%%%%%%%%%%%%%%
A well-known approach to \ac{BSS} of instantaneous mixtures is \ac{ICA} \cite{bell_information-maximization_1995}. \RevTwo{In} its original form \RevFour{\cite{hyvarinen_independent_2001}}\RevTwo{,} \ac{ICA} is not \Walter{directly} applicable to convolutive mixtures \RevTwo{captured by multiple sensors, due to the memory in the mixing system \RevThree{already} caused by relative delays between the signals captured \RevThree{by} different sensors}. \RevTwo{Therefore,} the source separation problem \RevTwo{is usually transformed} into the \Walter{\ac{STFT}} domain and \ac{ICA} \RevTwo{is applied} in each frequency bin independently, which is \Walter{commonly termed} \RevTwo{\acf{FD-ICA}} \RevThree{\cite{vincent_audio_2018}}. \Walter{For its representation}, we define the \Walter{microphone signal} vector at time-frequency bin $(\Freqix,\BlockIdx)$ as
\begin{equation}
\Freq{\x}(\Freqix,\BlockIdx) = \left[\Freq{x}_1(\Freqix,\BlockIdx),\dots,\Freq{x}_P(\Freqix,\BlockIdx)\right]\transp \in \mathbb{C}^P
\end{equation}
and the corresponding vector of demixed signals as
\begin{equation}
\Freq{\y}(\Freqix,\BlockIdx) = \left[\Freq{y}_1(\Freqix,\BlockIdx),\dots,\Freq{y}_P(\Freqix,\BlockIdx)\right]\transp \in \mathbb{C}^P.
\end{equation}
The demixing operation is applied \Walter{to} each frequency bin $\Freqix$ and time frame $\BlockIdx$ individually \Walter{and independently}
\begin{equation}
	\underbrace{\begin{bmatrix}
		\Freq{y}_1(\Freqix,\BlockIdx)\\
		\vdots\\
		\Freq{y}_P(\Freqix,\BlockIdx)
	\end{bmatrix}}_{\Freq{\y}(\Freqix,\BlockIdx)} = \underbrace{\begin{bmatrix}
	\Freq{w}_{11}(\Freqix)&\dots&\Freq{w}_{P1}(\Freqix)\\
	\vdots&\ddots&\vdots\\
	\Freq{w}_{1P}(\Freqix)&\dots&\Freq{w}_{PP}(\Freqix)
\end{bmatrix}}_{\WINST(\Freqix)}\underbrace{\begin{bmatrix}
\Freq{x}_1(\Freqix,\BlockIdx)\\
\vdots\\
\Freq{x}_P(\Freqix,\BlockIdx)
\end{bmatrix}}_{\Freq{\x}(\Freqix,\BlockIdx)}.
\label{eq:ICA_demixing}
\end{equation}
Note that $\WINST(\Freqix) \RevTwo{\in \mathbb{C}^{P\times P}}$ contains \Walter{generally complex-valued} scalar elements \RevFour{$\Freq{w}_{pq}(\Freqix)$, $p,q\in \{1,\dots,P\}$}, i.e., $\WINST(\Freqix)$ \Walter{represents} instantaneous demixing.
%Output $q$
%\begin{equation}
%	y_q(\tau) = \sum_{p=1}^{P} W_{qp}x_p(n)
%\end{equation}
The relation between the \ac{PDF} of the input signal vector $\Freq{\x}(\Freqix,\BlockIdx)$ and the output signal vector $\Freq{\y}(\Freqix,\BlockIdx)$ can be represented \Walter{as} a \Walter{linear mapping} of \RevTwo{complex-valued random} variables using \eqref{eq:ICA_demixing} \RevTwo{\cite{moreau_blind_2013}}
\begin{equation}
	p(\Freq{\y}(\Freqix,\BlockIdx)) = \frac{1}{\left\vert \det \WINST(\Freqix)\right\vert^2}\ p\left(\Freq{\x}(\Freqix,\BlockIdx)\right).
	\label{eq:PDF_transformFDICA}
\end{equation}
Using \eqref{eq:PDF_transformFDICA}, the cost function for frequency bin $\Freqix$ \Walter{reflecting} the mutual information can be written in terms of the Kullback-Leibler divergence \RevTwo{of the observed joint \ac{PDF} from the ideally separated and thus independent signals} as \Walter{\cite{cover_elements_2006}}
\begin{align}
	&\RevTwo{\tilde{J}}_\text{FD-ICA}(\WINST(\Freqix)) = \mathcal{KL} \left\lbrace p(\Freq{\y}(\Freqix,\BlockIdx))\bigg \vert \bigg \vert  \prod_{q=1}^{P} p(\Freq{y}_q(\Freqix,\BlockIdx))\right\rbrace\\[1ex]
	&=  -\underbrace{\mathcal{H}_{p(\Freq{\x}(\Freqix,\BlockIdx))}}_\text{const.} -\sum_{q=1}^{P} \mathcal{E}_{\RevFour{\text{frame}}}\left[\log p\left(\Freq{y}_q(\Freqix) \right)\right]\cdots\\
	&\qquad\qquad\qquad\qquad\qquad \cdots - 2\log \left\vert  \det\WINST(\Freqix) \right\vert,\notag
\end{align}
where $\mathcal{H}_{p(\Freq{\x}(\Freqix,\BlockIdx))}$ denotes the differential entropy of $p(\Freq{\x}(\Freqix,\BlockIdx))$, which \RevFour{does not depend on} the demixing matrix $\WINST(\Freqix)$. \RevFour{The expectation operator is denoted by $\mathcal{E}_{\RevFour{\text{frame}}}$, where the subscript emphasizes that a statistic over log likelihood values of signal frames has to be estimated.} Note that in \ac{FD-ICA} independent cost functions are used for each frequency bin. After neglecting the constant term and by approximating the expectation operator \RevFour{$\mathcal{E}_{\RevFour{\text{frame}}}[\cdot]$} \Walter{by arithmetic averaging over \RevFour{all available $\NumBlocks$ time frames}}, we obtain \RevThree{as} the \RevThree{simplified} \ac{FD-ICA} cost function \Walter{\cite{vincent_audio_2018}}
\begin{align}
\label{eq:costFunction_FDICA_approx}
	&J_\text{FD-ICA}(\WINST(\Freqix)) \ \RevTwo{:=}\\[1ex]
	&\quad -\frac{1}{\NumBlocks}\sum_{\BlockIdx=0}^{\NumBlocks-1}\left\lbrace \sum_{q=1}^{P}\log p(\Freq{y}_q(\Freqix,\BlockIdx))\right\rbrace - 2\log\left\vert \det \WINST(\Freqix) \right\vert. \notag
\end{align}
\RevFour{Minimizing the \ac{FD-ICA} cost function \eqref{eq:costFunction_FDICA_approx} decreases the differential entropy of the output signals in each frequency bin, i.e., decreases their Gaussianity, which is expressed by the first term in \eqref{eq:costFunction_FDICA_approx}. As a signal mixture is more Gaussian than its input signals, minimizing Gaussianity reverses the mixing process. The second term in \eqref{eq:costFunction_FDICA_approx} \RevFive{promotes} linearly independent demixing vectors, i.e., rows of the demixing matrix $\WINST(\Freqix)$, and, hence, ensures that different output signals are extracted. Also trivial solutions by decreasing the first term of \eqref{eq:costFunction_FDICA_approx} by scaling the output signals by demixing matrices comprising elements of small magnitudes are avoided by the second term in \eqref{eq:costFunction_FDICA_approx}. \RevFive{Please refer to \cite{brendel_diss} for a more detailed discussion of the two terms of the cost function in \eqref{eq:costFunction_FDICA_approx}.}}
%%%%%%%%%%%%%%%%%%%%%%%%%%%%%%%%%%%%%%%%%%%%%%%%%%%%%%%%%%%%%%%%%%%%%%%%%%%%
\section{Independent Vector Analysis}
\label{sec:IVA}
%%%%%%%%%%%%%%%%%%%%%%%%%%%%%%%%%%%%%%%%%%%%%%%%%%%%%%%%%%%%%%%%%%%%%%%%%%%%
\Walter{As a \RevTwo{particular method} to} address the \Walter{above-mentioned} internal permutation problem, \ac{IVA} \RevTwo{modifies} the cost function of \ac{FD-ICA} by incorporating statistical coupling \RevTwo{between} the frequency \RevTwo{bins} for the respective separated sources \RevTwo{as shown below}.

The demixing operation in \ac{IVA} is \RevTwo{still} applied for each frequency bin $\Freqix$ and time frame $\BlockIdx$ separately \RevTwo{according to \eqref{eq:ICA_demixing}.} In contrast to \ac{FD-ICA}, we now model \RevFour{the vector capturing all $\FreqixMax$ frequency bins of} the \RevTwo{\ac{STFT}-domain output vector for channel $q$} 
\begin{equation}
	\Freq{\y}_q(\BlockIdx) = \left[\Freq{y}_q(1,\BlockIdx),\dots,\Freq{y}_q(\FreqixMax,\BlockIdx)\right]\transp \in \mathbb{C}^\FreqixMax
\end{equation}
to follow a \RevTwo{$\FreqixMax$-variate} \ac{PDF} $p(\Freq{\y}_q\RevTwo{(\BlockIdx)})$ \Walter{\RevTwo{capturing the statistical} dependencies  between \RevTwo{all} frequency \RevTwo{bins to avoid} the internal permutation problem.} 

The relation between the \ac{PDF} of the input signal vectors of all frequency bins $\Freq{\x}(\Freqix,\BlockIdx)$, $\forall \Freqix$ and the output signal vectors $\Freq{\y}(\Freqix,\BlockIdx)$, $\forall \Freqix$ can \RevTwo{- analogously to \eqref{eq:PDF_transformFDICA} -} be represented by a \Walter{linear mapping} of \RevTwo{complex-valued random} variables using \eqref{eq:ICA_demixing} \RevTwo{\cite{moreau_blind_2013}}
\begin{align}
&p\left(\Freq{\y}(1,\BlockIdx),\dots,\Freq{\y}(\FreqixMax,\BlockIdx)\right) = \notag\\[1ex]
&\qquad\frac{1}{\left\vert \prod_{\Freqix = 1}^{\FreqixMax} \det \WINST(\Freqix)\right\vert^2}\ p\left(\Freq{\x}(1,\BlockIdx),\dots,\Freq{\x}(\FreqixMax,\BlockIdx)\right).\vspace{2pt}
\label{eq:PDF_transformIVA}
\end{align}
\RevThree{Note that the scaling term stems from the determinant of a $\FreqixMax Q \times\FreqixMax Q$ block diagonal matrix with $\WINST(\Freqix)$ on its block diagonal, see \eqref{eq:IVA_demixingComplete} and the illustration in Fig.~\ref{fig:resort_IVA_demixing_matrix} on the left.} The \ac{IVA} cost function \Walter{reflecting} minimum mutual information can \RevTwo{then again} be written in terms of the Kullback-Leibler divergence for frequency bin $\Freqix$ as
\begin{align}
&\RevTwo{\tilde{J}}_\text{IVA}\RevFour{\left(\WINST(1),\dots,\WINST(\FreqixMax)\right)} =\\
&=\mathcal{KL} \left\lbrace p(\Freq{\y}(1,\BlockIdx),\dots,\Freq{\y}(\FreqixMax,\BlockIdx))\bigg \vert \bigg \vert  \prod_{q=1}^{P} p(\Freq{\y}_q)\right\rbrace\notag\\[1ex]
&=  -\underbrace{\mathcal{H}_{p(\Freq{\x}(1,\BlockIdx),\dots,\Freq{\x}(\FreqixMax,\BlockIdx))}}_\text{const.} -\sum_{q=1}^{P} \mathcal{E}_{\RevFour{\text{frame}}}\left[\log p(\Freq{\y}_q)\right] \cdots\notag\\
& \qquad \qquad \qquad \qquad \cdots - 2\sum_{\Freqix=1}^{\FreqixMax}\log \left\vert  \det\WINST(\Freqix) \right\vert,\notag
\end{align}
where $\mathcal{H}_{p(\Freq{\x}(1,\BlockIdx),\dots,\Freq{\x}(\FreqixMax,\BlockIdx))}$ denotes the differential entropy of $p(\Freq{\x})$ which is constant w.r.t. the demixing \RevFour{matrices} $\WINST(\Freqix)$, $\Freqix\in\{1,\dots,\FreqixMax\}$ . After neglecting the constant term and by approximating the expectation operator \RevFour{$\mathcal{E}_{\RevFour{\text{frame}}}[\cdot]$} \Walter{by arithmetic averaging over \RevFour{all $\NumBlocks$ available time frames}}, we obtain the cost function~\Walter{\cite{vincent_audio_2018}}
\begin{align}
\label{eq:costFunction_IVA_approx}
&J_\text{IVA}\RevFour{\left(\WINST(1),\dots,\WINST(\FreqixMax)\right)} \ \RevTwo{:=}\\[1ex]
&\quad \sum_{q=1}^{P}\RevThree{\frac{1}{\NumBlocks}\sum_{\BlockIdx=0}^{\NumBlocks-1}}\left\lbrace-\log p(\Freq{\y}_q(\BlockIdx))\right\rbrace - 2\sum_{\Freqix=1}^{\FreqixMax}\log\left\vert \det \WINST(\Freqix) \right\vert \notag
\end{align}
\RevFour{depending on} all frequency bins $\Freqix$. 
%\RevTwo{Hereby, $\hat{\mathcal{E}}\{\cdot\} =  \frac{1}{\NumBlocks}\sum_{\BlockIdx=0}^{\NumBlocks-1}(\cdot)$ denotes a sample average \RevThree{over a time interval of $\NumBlocks$ time-frequency bins along the time axis for each channel $q$}.} 
\RevTwo{We note as the obvious difference to \ac{ICA} that \RevThree{by using a multivariate source prior,} the cost function for each frequency bin $\Freqix$ takes all other frequency bins of all output channels and the demixing systems into account.}

\RevFour{Considering \eqref{eq:costFunction_IVA_approx} and assuming statistical independence between individual frequency bins
\begin{equation}
	p(\Freq{\y}_q(\BlockIdx)) = \prod_{\Freqix = 1}^{\FreqixMax}p(\Freq{y}_q\left(\Freqix,\BlockIdx)\right),
\end{equation}
we directly obtain independent cost functions per frequency bin. This directly leads to \eqref{eq:costFunction_FDICA_approx} and confirms that \ac{FD-ICA} is a special case of \ac{IVA} \cite{vincent_audio_2018}.}
\begin{figure*}
	\begin{tikzpicture}[scale=0.95]
	%			\draw[dashed] (0,0)grid(11,6);
	\draw[thick] (0,0)--(0,6)--(2,6)--(2,0)--(0,0);
	\draw[thick] (1,0)--(1,6);
	\draw[thick] (0,3)--(2,3);
	
	\draw[thick,fill = lightgray] (0,6)--(1,5)--(1,3.5)--(0,4.5)--(0,6);
	\draw[thick, fill = lightgray] (1,6)--(2,5)--(2,3.5)--(1,4.5)--(1,6);
	\draw[thick, fill = lightgray] (0,3)--(1,2)--(1,0.5)--(0,1.5)--(0,3);
	\draw[thick,fill = lightgray] (1,3)--(2,2)--(2,0.5)--(1,1.5)--(1,3);
	
	\draw[<->](-0.2,3)--(-0.2,6);
	\node at (-0.5,4.5){$2L$};
	\draw[<->](0,6.2)--(1,6.2);
	\node at (0.5,6.5){$D$};
	\node at (0.5,4.7){$\mathbf{W}_{11}$};
	\node at (1.5,4.7){$\mathbf{W}_{12}$};
	\node at (0.5,1.7){$\mathbf{W}_{21}$};
	\node at (1.5,1.7){$\mathbf{W}_{22}$};
	
	\node at (1,-0.5){$\W$, see \eqref{eq:TRINICON_demixingModel}};
	
	\draw[->,thick] (2.5,3)--(3.5,3);

	\begin{scope}[shift = {(-1,0)}]
	\draw[thick] (5,0)--(5,6)--(11,6)--(11,0)--(5,0);
	\draw[thick] (8,0)--(8,6);
	\draw[thick] (5,3)--(11,3);
	
	\draw[thick,fill = lightgray] (5,6)--(6,5)--(6,3.5)--(5,4.5)--(5,6);
	\draw[thick, fill = lightgray] (8,6)--(9,5)--(9,3.5)--(8,4.5)--(8,6);
	\draw[thick, fill = lightgray] (5,3)--(6,2)--(6,0.5)--(5,1.5)--(5,3);
	\draw[thick,fill = lightgray] (8,3)--(9,2)--(9,0.5)--(8,1.5)--(8,3);
	
	\draw[thick] (6,5)--(8,3);
	\draw[thick] (9,2)--(11,0);
	\draw[thick] (6,6)--(6,0);
	\draw[thick] (9,6)--(9,0);
	\draw[thick] (6,5)--(8,5);
	\draw[thick] (9,2)--(11,2);
	
%	\draw[<->](4.8,3)--(4.8,6);
%	\node at (4.5,4.5){$2L$};
%	\draw[<->](5,6.2)--(6,6.2);
%	\node at (5.5,6.5){$D$};
%	\draw[<->](6,6.2)--(8,6.2);
%	\node at (7,6.5){$2L-D$};
%	\draw[<->](11.2,3)--(11.2,5);
%	\node at (12,4){$2L-D$};
	\node at (5.5,4.7){$\mathbf{W}_{11}$};
	\node at (8.5,4.7){$\mathbf{W}_{12}$};
	\node at (5.5,1.7){$\mathbf{W}_{21}$};
	\node at (8.5,1.7){$\mathbf{W}_{22}$};
	\node at (10,4.5){$\mathbf{0}_{2L\times(2L-D)}$};
	\node at (7,1.5){$\mathbf{0}_{2L\times(2L-D)}$};
	\node at (10,2.5){$\mathbf{0}_{D\times(2L-D)}$};
	\node at (7,5.5){$\mathbf{0}_{D\times(2L-D)}$};
	
	\node[fill = white] at (10,1){$\mathbf{I}_{2L-D}$};
	\node[fill = white] at (7,4){$\mathbf{I}_{2L-D}$};
	
	\node at (8,-0.5){$\tilde{\W}$, see \eqref{eq:TRINICON_extendedDemixingMat}};
	
%	\node at (11,-0.5){$\mathbf{I}_{(2L-D)\times(2L-D)}$};
	\end{scope}
	
	\begin{scope}[shift = {(7,0)}]
	\draw[thick] (5,0)--(5,6)--(11,6)--(11,0)--(5,0);
	\draw[thick] (7,0)--(7,6);
	\draw[thick] (5,4)--(11,4);
	
	\draw[thick,fill = lightgray] (5,6)--(6,5)--(5,5)--(5,6);
	\draw[thick,fill = lightgray] (6,6)--(7,5)--(6,5)--(6,6);
	\draw[thick,fill = lightgray] (5,5)--(6,4)--(5,4)--(5,5);
	\draw[thick,fill = lightgray] (6,5)--(7,4)--(6,4)--(6,5);
	
	\draw[thick,fill = lightgray] (5,4)--(6,4)--(6,2.5)--(5,3.5)--(5,4);
	\draw[thick,fill = lightgray] (6,4)--(7,4)--(7,2.5)--(6,3.5)--(6,4);
	\draw[thick,fill = lightgray] (5,2)--(6,2)--(6,0.5)--(5,1.5)--(5,2);
	\draw[thick,fill = lightgray] (6,2)--(7,2)--(7,0.5)--(6,1.5)--(6,2);
	
	\draw[thick] (6,0)--(6,6);
	
	% arrow to right figure
	\draw[->,thick] (3.5,3)--(4.5,3);
	
%	\node at (10,3){$\mathbf{0}$};
%	\node at (8,1){$\mathbf{0}$};
%	\draw[thick] (5,6)--(11,0);
%	\draw[thick] (5,2)--(11,2);
%	\draw[thick] (9,0)--(9,4);
	
	\node at (9,5){$\mathbf{0}_{2D\times(4L-2D)}$};
	\node at (9,2){$\mathbf{I}_{4L-2D}$};
	
	\draw[<->](4.8,4)--(4.8,6);
	\node at (4.3,5){$2D$};
	\draw[<->](5,6.2)--(7,6.2);
	\node at (6,6.5){$2D$};
	\draw[<->](4.8,0)--(4.8,2);
	\node at (4,1){$2L-D$};
	
	\node at (8,-0.5){$\bar{\W}$, see \eqref{eq:TRINICON_perm_ext_demMat}};
	
	\end{scope}
	\end{tikzpicture}
	\caption{Extension of the demixing filter matrix $\W$ of \ac{TRINICON} for the \RevThree{case} $P=2$ (cf. \cite[Figure B.1]{aichner_acoustic_2007}). On the left, the Toeplitz matrix $\W$ is depicted, which is extended to the matrix $\tilde{\W}$ by inserting identity matrices $\mathbf{I}_{2L-D}$ and all-zero matrices, which is shown in the middle. Permutation of rows and columns yields the matrix on the right, which is used to calculate the determinant of the extended demixing filter matrix $\tilde{\W}$.}
	\label{fig:extended_demixingMatrix}
\end{figure*}
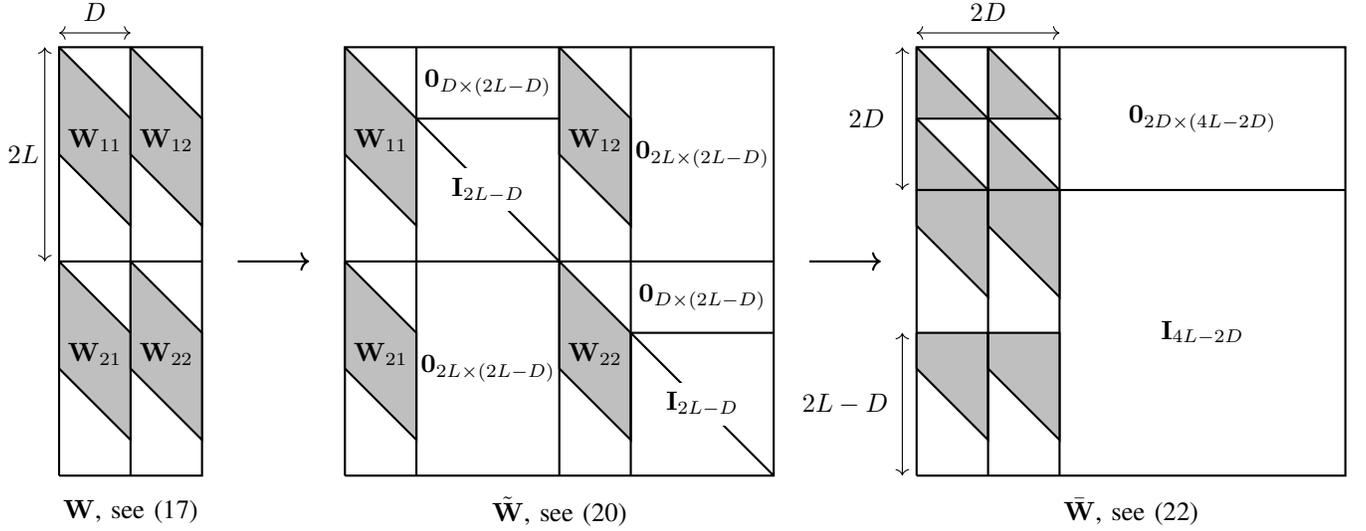
%%%%%%%%%%%%%%%%%%%%%%%%%%%%%%%%%%%%%%%%%%%%%%%%%%%%%%%%%%%%%%%%%%%%%%%%%%%%
\section{TRINICON}
\label{sec:TRINICON}
%%%%%%%%%%%%%%%%%%%%%%%%%%%%%%%%%%%%%%%%%%%%%%%%%%%%%%%%%%%%%%%%%%%%%%%%%%%%
\RevTwo{As an alternative approach to broadband convolutive \ac{BSS}, we briefly review the generic \ac{TRINICON} framework for blind \ac{MIMO} signal processing, before we establish the link to \ac{FD-ICA} and \ac{IVA}.} The \ac{TRINICON} cost function is, in contrast to the previously discussed approaches, defined in \RevTwo{the} \Walter{time domain}. Therefore, we define the time-domain microphone signal vector \RevThree{with block processing rate $R$, i.e., relative shift between two processed frames\footnote{\RevFour{Note that $\mathbf{x}_p(\TriBlockIdx)$ contains $2L$ samples instead of $L+D-1$ samples required for linear convolution due to consistency with the literature, where this choice is made for deriving concise \ac{STFT}-domain formulations \cite{aichner_acoustic_2007}.}},}
\begin{equation}
	\mathbf{x}_p(\TriBlockIdx) = \left[x_p(\TriBlockIdx \RevThree{R}),\dots, x_p(\TriBlockIdx \RevThree{R}-2L+1)\right]^\text{T} \in \mathbb{R}^{2L}
	\label{eq:TRI_def_x}
\end{equation}
 and the time-domain vector of demixed signals
\begin{equation}
	\mathbf{y}_{\RevTwo{q}}(\TriBlockIdx) = \left[y_q(\TriBlockIdx \RevThree{R}),\dots, y_q(\TriBlockIdx \RevThree{R}-D+1)\right]^\text{T}  \in \mathbb{R}^D.
\end{equation}
The relation between the microphone signals and the demixed signals is modeled by a linear convolution
\begin{equation}
	\underbrace{\begin{bmatrix}
		\mathbf{y}_1(\TriBlockIdx)\\
		\vdots\\
		\mathbf{y}_P(\TriBlockIdx)
	\end{bmatrix}}_{\mathbf{y}(\TriBlockIdx)} = \underbrace{\begin{bmatrix}
	\W_{11}&\dots&\W_{1P}\\
	\vdots&\ddots&\vdots\\
	\W_{P1}&\dots&\W_{PP}
\end{bmatrix}^\text{T}}_{\W\transp}\underbrace{\begin{bmatrix}
\mathbf{x}_1(\TriBlockIdx)\\
\vdots\\
\mathbf{x}_P(\TriBlockIdx)
\end{bmatrix}}_{\mathbf{x}(\TriBlockIdx)}
\label{eq:TRINICON_demixingModel}
\end{equation}
with the \Walter{$2L\times D$} convolution matrix \RevTwo{of \RevThree{Toeplitz} structure \cite{aichner_acoustic_2007}}\vspace{2pt}
\begin{equation}
	\W_{pq} = \begin{bmatrix}
	w_{pq,0}&0&\cdots&0\\
	w_{pq,1}&w_{pq,0}&\ddots&\vdots\\
	\vdots&w_{pq,1}&\ddots&0\\
	w_{pq,L-1}&\vdots&\ddots&\Walter{w_{pq,0}}\\
	0&w_{pq,L-1}&\ddots&w_{pq,1}\\
	\vdots&&\ddots&\vdots\\
	0&\cdots&0&w_{pq,L-1}\\
	0&\cdots&0&0\\
	\vdots&\cdots&\vdots&\vdots\\
	0&\cdots&0&0\\
	\end{bmatrix}.\vspace{2pt}
	\label{eq:convolution_matrix}
\end{equation}
To show the similarities in the derivation compared \RevFour{to} \RevThree{\ac{FD-ICA} and \ac{IVA}}, we will discuss the crucial steps in the derivation of the \ac{TRINICON} cost function \cite{aichner_acoustic_2007,buchner_broadband_2010}. To this end, we have to perform a change of variables \RevFour{corresponding to} \eqref{eq:PDF_transformFDICA} and \eqref{eq:PDF_transformIVA}, which includes the calculation of the determinant of the demixing matrix. Unfortunately, $\mathbf{W}$ is not quadratic as the submatrices $\W_{pq}$ are not quadratic\RevTwo{, so that the determinant does not exist (see Fig.~\ref{fig:extended_demixingMatrix}, left)}. \RevFour{Therefore, to facilitate} the change of variables, we append the last $2L-D$ elements of the microphone signals, denoted as $\tilde{\x}_j(\TriBlockIdx)$, to the demixed signals and extend the demixing filter matrix to $\tilde{\W}$ \RevTwo{(see Fig.~\ref{fig:extended_demixingMatrix}, center)}, which yields\vspace{2pt}
\begin{equation}
	\left[\y_1\transp(\TriBlockIdx),\tilde{\x}_1\transp(\TriBlockIdx),\dots,\y_P\transp(\TriBlockIdx),\tilde{\x}_P\transp(\TriBlockIdx)	\right]\transp = \tilde{\W}\transp \x(\TriBlockIdx).\vspace{2pt}
\end{equation}
The extended demixing filter matrix yields correspondingly
\begin{gather}
	\tilde{\W} = \begin{bmatrix}
	\mathbf{W}_{11}&\begin{bmatrix}
	\mathbf{0}_{D\times 2L-D}\\\mathbf{I}_{2L-D}
	\end{bmatrix}&\dots&\mathbf{W}_{1P}&\mathbf{0}_{2L\times 2L-D}\\
	\vdots&&\ddots&&\vdots\\
	\mathbf{W}_{P1}&\mathbf{0}_{2L\times 2L-D}&\dots&\mathbf{W}_{PP}&\begin{bmatrix}
	\mathbf{0}_{D\times 2L-D}\\\mathbf{I}_{2L-D}
	\end{bmatrix}
	\end{bmatrix}. \raisetag{-7pt}
	\label{eq:TRINICON_extendedDemixingMat}
\end{gather}
\RevTwo{In the following, we assume that $\tilde{\W}$ has full rank.}
Now, we use the rule \RevThree{for a linear mapping of} real-valued random vectors of different length \cite{papoulis_probability_1984}, i.e., we perform a \Walter{linear mapping} of variables \RevThree{by \eqref{eq:TRINICON_extendedDemixingMat}} with subsequent marginalization
\begin{align}
\label{eq:marginalization}
	&p (\y(\TriBlockIdx)) =\\[1ex]
	& \quad \frac{1}{\vert \det \tilde{\W}\vert}\int_{-\infty}^{\infty}\cdots\int_{-\infty}^{\infty} p \left(\x(\TriBlockIdx)\right) \mathrm{d}\tilde{\x}_1(\TriBlockIdx)  \cdots \mathrm{d}\tilde{\x}_P(\TriBlockIdx).\notag
\end{align}
Note that we can use in this case a transformation rule for real-valued random vectors as we operate in time domain compared to \ac{FD-ICA} and \ac{IVA}, which operate in \RevThree{the} \RevTwo{\ac{STFT}} domain.

We rearrange the rows and columns of the extended demixing matrix $\tilde{\W}$ by applying the permutation matrices $\mathbf{P}_1$ and $\mathbf{P}_2$, to obtain the following permuted extended demixing matrix \RevTwo{(see Fig.~\ref{fig:extended_demixingMatrix}, right)}
\begin{gather}
	\bar{\W} = \mathbf{P}_1 \tilde{\W} \mathbf{P}_2= \begin{bmatrix}
	\left(\Window_{2PL\times PD}^{1_D0}\right)\transp\mathbf{W}& \mathbf{0}_{PD\times(2PL-PD)}\\[1.5ex]
	\left(\Window_{2PL\times PD}^{01_{2L-D}}\right)\transp\mathbf{W}&\mathbf{I}_{2PL-PD}
	\end{bmatrix}.\raisetag{-3pt}
	\label{eq:TRINICON_perm_ext_demMat}
\end{gather} 
Hereby, we defined the following window matrices using the Kronecker product $\otimes$
\begin{equation}
\DFTWindowPLD = \mathbf{I}_P \otimes \begin{bmatrix}\mathbf{I}_D\\ \mathbf{0}_{(2L-D)\times D}\end{bmatrix}
\end{equation}
and
\begin{equation}
\Window_{2PL\times PD}^{01_{2L-D}} = \mathbf{I}_P \otimes \begin{bmatrix}\mathbf{0}_{D\times(2L-D)}\\ \mathbf{I}_{(2L-D)}\end{bmatrix}
\end{equation}
The application of the permutation matrices does not change the absolute value of the determinant of $\tilde{\W}$, i.e.,\vspace{2pt}
\begin{equation}
	\label{eq:determinant_permutationMat}
%	\vert \det\tilde{\W}\vert &=  \vert\det\mathbf{P}_1\vert\,\vert \det\tilde{\W}\vert\,\vert\det\mathbf{P}_2\vert \\
%	&=  \vert\det\mathbf{P}_1\tilde{\W}\mathbf{P}_2\vert= \vert \det\bar{\W}\vert.\notag
\RevTwo{\left\vert \det\tilde{\W}\right\vert =  \left\vert\det\mathbf{P}_1\right\vert\,\left\vert \det\tilde{\W}\right\vert\,\left\vert\det\mathbf{P}_2\right\vert = \left\vert \det\bar{\W}\right\vert,}\vspace{2pt}
\end{equation}
\RevFour{due to \RevFive{$\vert\det\mathbf{P}_1\vert = \vert\det\mathbf{P}_2\vert = 1$} \cite{harville_matrix_2008}.} Furthermore, we observe that $\bar{\W}$ is a lower block-triangular matrix. Hence, its determinant is given as \cite{harville_matrix_2008}
\begin{align}
	\left\vert \det\bar{\W}\right\vert &= \left\vert \det\left(\Window_{2PL\times PD}^{1_D0}\right)\transp\mathbf{W}\right\vert  \left\vert \det\mathbf{I}_{2PL-PD}\right\vert\\[1ex]
	&= \left\vert \det\left(\Window_{2PL\times PD}^{1_D0}\right)\transp\mathbf{W}\right\vert. \notag
\end{align}
\RevTwo{Thus}, we can conclude that the determinants of the extended demixing matrix $\tilde{\W}$ and a channel-wise truncated version of the Toeplitz matrices $(\Window_{2PL\times PD}^{1_D0})\transp\mathbf{W}$ are identical (see Fig.~\ref{fig:extended_demixingMatrix}) and by evaluating the integral \eqref{eq:marginalization} \RevTwo{we finally arrive} at
\begin{equation}
	p(\mathbf{y}(\TriBlockIdx)) = \frac{1}{\left\vert \det \left(\Window_{2PL\times PD}^{1_D0}\right)\transp\mathbf{W}\right\vert}\ p(\bar{\mathbf{x}}(\TriBlockIdx)),
	\label{eq:TRINICON_changeOfVariables}
\end{equation}
with the \RevTwo{shortened} vector of microphone signals
\begin{equation}
	\bar{\mathbf{x}}(\TriBlockIdx) = \left[\bar{\mathbf{x}}_1^\text{T}(\TriBlockIdx),\dots,\bar{\mathbf{x}}_P^\text{T}(\TriBlockIdx)\right]^\text{T} \in \mathbb{R}^{PD},
\end{equation}
where each element is defined as
\begin{equation}
	\bar{\mathbf{x}}_p(\TriBlockIdx) = \left[x_p(\TriBlockIdx \RevThree{R}),\dots,x_p(\TriBlockIdx \RevThree{R}-D\RevTwo{+1})\right]^\text{T} \in \mathbb{R}^D
\end{equation}
\RevTwo{and thus comprises only $D<2L$ input samples rather than $2L$ as in \eqref{eq:TRI_def_x}. This is due to the marginalization in \eqref{eq:marginalization}, i.e., the influence of the microphone signal components $\tilde{\x}_p(\TriBlockIdx)$ is absorbed in the \ac{PDF} $p(\bar{\mathbf{x}}(\TriBlockIdx))$.} \RevFour{Hence, for the statistical properties of the demixed signals expressed by the \ac{PDF} $p(\mathbf{y}(\TriBlockIdx))$ only a dependence on the $D$ most recent input samples $\bar{\mathbf{x}}(\TriBlockIdx)$ \RevFive{is modeled by \eqref{eq:TRINICON_changeOfVariables}}. \RevFive{Thereby}, the temporal context of the demixed signals and the observed signals is chosen to be equal. \RevFive{Note} that for this line of arguments, only the statistical properties of observed signals and demixed signals are related \RevFive{to} each other \RevFive{not the signals themselves}, i.e., by applying the demixing matrix as in \eqref{eq:TRINICON_demixingModel} the \RevFive{actual} vector of demixed signals $\mathbf{y}(\TriBlockIdx)\in\mathbb{R}^{PD}$ depends on all $2LP$ samples of the observed signal vector $\mathbf{x}(\TriBlockIdx)$.} 
%In the following, we use the substitution $\TriBlockIdx = \TriniconBlock L + j$, when appropriate, for the sake of a concise notation and introduce the mapping 
%\begin{equation}
%	\BlockIdx = f(\TriBlockIdx)
%	\label{eq:mapping_BlockFrame}
%\end{equation} 
%from the sample index $\TriBlockIdx$ to the \RevTwo{\ac{STFT} frame} index $\BlockIdx$, which enumerates the signal blocks. \RevTwo{This relation will be used in the comparison of the methods in Sec.~\ref{sec:relation_BSSAlgorithms}}.

The \ac{TRINICON} \Walter{BSS} cost function is defined \RevTwo{as \cite{aichner_acoustic_2007}}
\begin{gather}
	\RevTwo{\check{J}_{\text{TRI}}(\TriBlockIdx',\mathbf{W}) = \sum_{\TriBlockIdx=0}^{\infty}\beta(\TriBlockIdx,\TriBlockIdx')\mathcal{KL}\left\lbrace p(\mathbf{y}(\TriBlockIdx)) \bigg\Vert\prod_{q=1}^{P}p(\mathbf{y}_q(\TriBlockIdx)) \right\rbrace,}\raisetag{-3pt}
	\label{eq:TRINICON_costFunction}
\end{gather}
where $\beta(\TriBlockIdx,\TriBlockIdx')$ is a weighting function controlling the influence of block \RevThree{$\TriBlockIdx$} on the cost function evaluated for block $\TriBlockIdx'$ and allowing for different \RevTwo{versions} of the algorithm. The second part of the cost function is the Kullback-Leibler divergence.
%, which can be approximated \Walter{by arithmetic averaging in the same sense as in \eqref{eq:costFunction_FDICA_approx} and \eqref{eq:costFunction_IVA_approx}} \cite{buchner_trinicon:_2004,buchner_blind_2004}
%\begin{equation}
%J_{\text{TRI}}(\TriniconBlock',\mathbf{W}) = \sum_{\TriniconBlock=0}^{\infty}\beta(\TriniconBlock,\TriniconBlock')\frac{1}{N}\sum_{j=0}^{N-1}\left\lbrace \log \frac{p(\mathbf{y}(\TriniconBlock L+j))}{\prod_{q=1}^{P}p(\mathbf{y}_q(\TriniconBlock L+j))} \right\rbrace.
%\end{equation}

The previously discussed algorithms are offline algorithms, i.e., they operate on the \RevFour{complete} recorded signal. Therefore, we choose the weighting function $\beta(\TriBlockIdx,\TriBlockIdx')$ to \Walter{represent} the offline version of the \ac{TRINICON} \Walter{\ac{BSS}} cost function \cite[Section~3.5.1]{aichner_acoustic_2007} for comparison
\begin{equation}
\RevTwo{\beta(\TriBlockIdx,\TriBlockIdx') := \left\lbrace\begin{matrix}
\frac{1}{\NumBlocks}&\mathrm{for}\ \TriBlockIdx'\in\{1,\dots,\NumBlocks\}\\
0&\mathrm{otherwise}
\end{matrix}\right. .}
\end{equation}
Insertion of this choice of $\beta$ into the \ac{TRINICON} cost function \eqref{eq:TRINICON_costFunction} yields
\begin{align}
&\RevTwo{\tilde{J}_{\text{TRI}}(\mathbf{W}) =}
% \sum_{\TriniconBlock=0}^{\infty}\frac{1}{\NumBlocks}\epsilon_{0,(\NumBlocks-1)}(\TriniconBlock)\tilde{J}(\TriniconBlock,\mathbf{W})\notag\\
%&\quad = 
\RevTwo{\frac{1}{\NumBlocks}\sum_{\TriBlockIdx=0}^{\NumBlocks-1}\mathcal{KL}\left\lbrace p(\mathbf{y}(\TriBlockIdx)) \bigg\Vert\prod_{q=1}^{P}p(\mathbf{y}_q(\TriBlockIdx)) \right\rbrace} \notag\\[1ex]
&\RevTwo{\quad =  -\underbrace{\mathcal{H}_{p(\bar{\x}(1),\dots,\bar{\x}(\NumBlocks))}}_\text{const.} -\frac{1}{\NumBlocks}\sum_{\TriBlockIdx=0}^{\NumBlocks-1}\sum_{q=1}^{P} \mathcal{E}_{\RevFour{\text{sample}}}\left[\log p(\mathbf{y}_q(\TriBlockIdx))\right]}\notag\\[1ex]
&\quad \qquad \dots - \log \left\vert  \det\left(\DFTWindowPLD\right)\transp\mathbf{W}\right\vert, \label{eq:trinicon_costFunctionKL}
\end{align}
where we used \eqref{eq:TRINICON_changeOfVariables} for the change of variables. Note that, \RevFour{analogously to the cost function of \ac{FD-ICA} \eqref{eq:costFunction_FDICA_approx} and \ac{IVA} \eqref{eq:costFunction_IVA_approx} an \RevFive{arithmetic} average over all available $\NumBlocks$ signal blocks is introduced in \eqref{eq:trinicon_costFunctionKL} but also an additional expectation $\mathcal{E}_{\RevFour{\text{sample}}}[\cdot]$ over the log likelihood \RevFive{for} subsequent samples of the demixed signals is introduced. \RevFive{Note also} that the likelihood functions for \ac{FD-ICA}, \ac{IVA} and \ac{TRINICON} are different in general.}

The cost function with approximation of the Kullback-Leibler divergence \RevThree{by sample-wise averaging along the time axis for each channel $q$} \Walter{according to \eqref{eq:TRINICON_costFunction}}, \Walter{including} offline weight and neglecting constant terms yields
\begin{align}
	J_{\text{TRI}}(\mathbf{W}) &\ \RevTwo{:=} \RevTwo{\frac{1}{\NumBlocks}\sum_{\TriBlockIdx=0}^{\NumBlocks-1} \sum_{q=1}^{P}\hat{\mathcal{E}}_{\RevFour{\text{sample}}}\left\lbrace-\log p\left(\mathbf{y}_q(\TriBlockIdx)\right)\right\rbrace\dots} \notag\\[1ex]
	& \qquad \qquad +\log\left\vert \det \left(\DFTWindowPLD\right)\transp\mathbf{W} \right\vert.
	\label{eq:TRINICON_costFct_woKL}
\end{align}
The last term of \eqref{eq:TRINICON_costFct_woKL} can be written explicitly as
\begin{align}
	&\left(\RevTwo{\DFTWindowPLD}\right)\transp \W = \left(\mathbf{I}_P \otimes \left(\DFTWindowLD\right)\transp\right) \W\\[1.5ex]
	&= \begin{bmatrix}
	\left(\DFTWindowLD\right)\transp\mathbf{W}_{11}&\dots&\left(\DFTWindowLD\right)\transp\mathbf{W}_{1P}\notag\\
	\vdots&\ddots&\vdots\\
	\left(\DFTWindowLD\right)\transp\mathbf{W}_{P1}&\dots&\left(\DFTWindowLD\right)\transp\mathbf{W}_{PP}
	\end{bmatrix}.
\end{align}
\RevFive{Hence, the matrices $\DFTWindowLD$ extract the upper parts of the demixing filter matrices $\mathbf{W}_{pq}$ (see also Fig.~\ref{fig:extended_demixingMatrix} for an illustration). A detailed analysis of the different influence of the log det regularization terms of \ac{IVA} and \ac{TRINICON} can be found in \cite{brendel_diss}.}
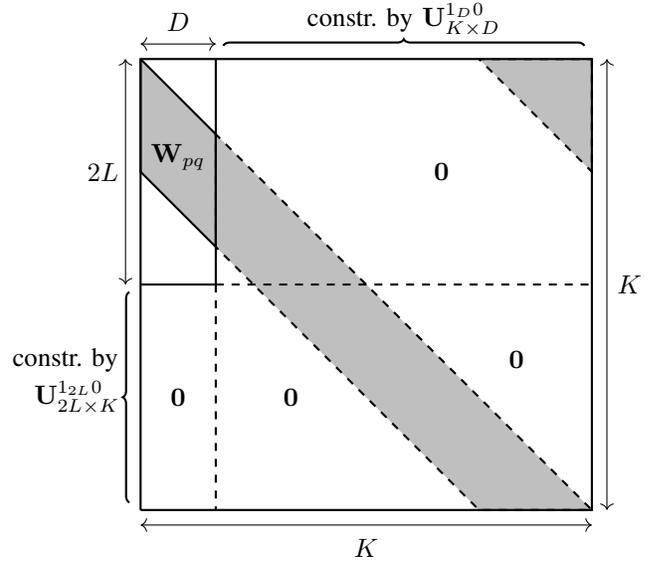
\begin{figure}
	\begin{tikzpicture}	
		\draw[thick,fill = lightgray] (5,6)--(6,5)--(6,3.5)--(5,4.5)--(5,6);
		\draw[dashed,thick,fill = lightgray] (6,3.5)--(6,5)--(11,0)--(9.5,0)--(6,3.5);
		\draw[dashed,thick,fill = lightgray] (11,6)--(11,4.5)--(9.5,6)--(11,6);
			
		\draw[thick] (5,0)--(5,6)--(11,6)--(11,0)--(5,0);
		\draw[thick] (5,3)--(6,3);
		\draw[thick,dashed] (6,3)--(11,3);
		\draw[thick] (6,6)--(6,3);
		\draw[thick,dashed] (6,3)--(6,0);	
			
		\draw[<->](4.8,3)--(4.8,6);
		\node at (4.5,4.5){$2L$};
		\draw[thick,decoration={brace,mirror,raise=-10pt},decorate](4.5,2.9) -- node[left=-10pt] {$\Window_{2L\times \DFTLength}^{1_{2L}0}$} (4.5,0.1);
		\node at (4,2)(){constr. by};
		\draw[<->](5,6.2)--(6,6.2);
		\node at (5.5,6.5){$D$};
		\draw[<->](5,-0.2)--(11,-0.2);
		\node at (8,-0.5){$\DFTLength$};
		\draw[thick,decoration={brace,mirror,raise=5pt},decorate](10.9,6) -- node[above=5pt] {constr. by $\Window_{\DFTLength\times D}^{1_{D}0}$} (6.1,6);
		\draw[<->](11.2,0)--(11.2,6);
		\node at (11.5,3){$\DFTLength$};
		\node at (5.5,4.7){$\W_{pq}$};
		\node at (9,4.5){$\mathbf{0}$};
		\node at (7,1.5){$\mathbf{0}$};
		\node at (5.5,1.5){$\mathbf{0}$};
		\node at (10,2){$\mathbf{0}$};
\end{tikzpicture}
	\caption{Extension of a subblock $\W_{pq}$ of the demixing filter matrix $\W$ to a circular matrix $\Wcirc$. This is accomplished by adding repeating elements until the matrix is a square matrix. Window matrices constrain the obtained circular matrix such that the original Toeplitz matrix can be reconstructed.}
	\label{fig:extend_demixingMatrix2Circular}
\end{figure}
\begin{figure*}
	\setcounter{MYtempeqncnt}{\value{equation}}
	\setcounter{equation}{43}
	\begin{align}
	\RevTwo{J_{\text{TRI-FD}}(\underline{\mathbf{W}})} &= \frac{1}{\NumBlocks}\sum_{\RevTwo{\TriBlockIdx}=0}^{\NumBlocks-1} \sum_{q=1}^{P}\RevTwo{\hat{\mathcal{E}}}_{\RevFour{\text{sample}}}\left\lbrace-\log p\left(\DFTWindowDR\FourierMatInv\Freq{\W}_{q}\Freq{\mathbf{x}}(\RevTwo{\TriBlockIdx})\right)\right\rbrace\notag\\[1ex]
	&\quad +\log\left\vert \det \begin{bmatrix}
	\DFTWindowDR\FourierMat \Freq{\W}_{11} \FourierMatInv \left(\DFTWindowDR\right)\transp&\dots&\DFTWindowDR\FourierMat \Freq{\W}_{1P} \FourierMatInv \left(\DFTWindowDR\right)\transp\\
	\vdots&\ddots&\vdots\\
	\DFTWindowDR\FourierMat \Freq{\W}_{P1} \FourierMatInv \left(\DFTWindowDR\right)\transp&\dots&\DFTWindowDR\FourierMat \Freq{\W}_{PP} \FourierMatInv \left(\DFTWindowDR\right)\transp
	\end{bmatrix} \right\vert \label{eq:TRINICON_costFct_DFT}\\[1ex]
	&= \frac{1}{\NumBlocks}\sum_{\RevTwo{\TriBlockIdx}=0}^{\NumBlocks-1}\RevTwo{\hat{\mathcal{E}}}_{\RevFour{\text{sample}}}\left\lbrace-\log p\left(\DFTWindowDR\FourierMatInv\Freq{\W}_{q}\Freq{\mathbf{x}}(\RevTwo{\TriBlockIdx})\right)\right\rbrace +\log\left\vert \det \left(\mathbf{I}_P \otimes \DFTWindowDR\FourierMat \right)  \Freq{\W} \left(\mathbf{I}_P \otimes \FourierMatInv\left(\DFTWindowDR\right)\transp \right) \right\vert \notag
	\end{align}
	\setcounter{equation}{\value{MYtempeqncnt}}
	\hrule
\end{figure*}
%%%%%%%%%%%%%%%%%%%%%%%%%%%%%%%%%%%%%%%%%%%%%%%%%%%%%%%%%%%%%%%%%%%%%%%%%%%%
\section{Relation\RevThree{s} between BSS Algorithms}
\label{sec:relation_between_BSS_algorithms}
%%%%%%%%%%%%%%%%%%%%%%%%%%%%%%%%%%%%%%%%%%%%%%%%%%%%%%%%%%%%%%%%%%%%%%%%%%%%
\Walter{In the previous sections, we \RevTwo{demonstrated} that the \RevTwo{cost functions} of the considered \ac{BSS} algorithms \ac{FD-ICA}, \ac{IVA} and \ac{TRINICON} follow \RevFour{comparable} ideas and \RevTwo{we exposed} \RevThree{different models underlying the different versions of the Kullback-Leibler divergence as the general cost function}.}

\RevThree{For a deeper insight into the underlying model assumptions, we} want to \RevTwo{investigate} the relation\RevThree{s} between the \ac{BSS} algorithms \RevFour{more closely} \RevThree{in the following}. To this end, we reformulate the \ac{TRINICON} and the \ac{IVA} cost \RevTwo{functions, respectively,} to obtain \RevTwo{representations} which allow their \Walter{analytical} comparison.
\subsection{Reformulation of TRINICON}
As \ac{FD-ICA} and \ac{IVA} are defined in the \RevTwo{\ac{STFT}} domain, we will formulate the \ac{TRINICON} cost function equivalently in the \RevTwo{\ac{STFT}} domain.

\RevTwo{First}, we expand the Toeplitz matrix $\W_{pq}\transp \in \mathbb{R}^{D\times 2L}$ to a circulant matrix $\Wcirc\in \mathbb{R}^{\DFTLength\times \DFTLength}$ with $\DFTLength\ \RevThree{\geq}\ 2L$ and \mbox{$L\RevFour{\geq} D$} by appending repeating elements to the Toeplitz matrix\Walter{, see also \cite{buchner_generalization_2005} and Fig.~\ref{fig:extend_demixingMatrix2Circular}}
\begin{equation}
	\W_{pq}\transp = \DFTWindowDR \Wcirc \DFTWindowRL
	\label{eq:windowed_circularMat}
\end{equation}
where the window matrices
\begin{equation}
\DFTWindowDR = \begin{bmatrix} \mathbf{I}_{D} & \mathbf{0}_{D\times (\DFTLength-D)}
\end{bmatrix}
\end{equation}
and
\begin{equation}
\DFTWindowRL = \begin{bmatrix} \mathbf{I}_{2L}\\ \mathbf{0}_{(\DFTLength-2L)\times 2L} 
\end{bmatrix}
\end{equation}
cut out the Toeplitz part from the circulant matrix. \RevFour{Next}, we use the property that a \Walter{\ac{DFT}}~matrix 
\begin{gather}
	\FourierMat = \frac{1}{\sqrt{\DFTLength}}\begin{bmatrix}
	1     & 1          & 1 & \cdots & 1\\
	1     & \RootUnity & \RootUnity^2 & \cdots & \RootUnity^{\DFTLength-1} \\
	1     & \RootUnity^2 & \RootUnity^4 & \cdots & \RootUnity^{2(\DFTLength-1)} \\
	\vdots&\vdots&\vdots&\ddots&\vdots\\
	1     & \RootUnity^{\DFTLength-1} & \RootUnity^{2(\DFTLength-1)} & \cdots & \RootUnity^{(\DFTLength-1)(\DFTLength-1)}
	\end{bmatrix},\raisetag{-3pt}
\end{gather}
with $\RootUnity = \exp\left(\frac{-j2\pi}{\DFTLength}\right)$, diagonalizes \Walter{any} circulant matrix of corresponding dimensions \cite{golub_matrix_1996}
\begin{equation}
\Freq{\W}_{pq} = \FourierMat \Wcirc \FourierMatInv
\end{equation}
or equivalently
\begin{equation}
	\Wcirc = \FourierMatInv \Freq{\W}_{pq} \FourierMat.\vspace{2pt}
	\label{eq:circMat_fromFreqW}
\end{equation}
\Walter{Hereby, the diagonal matrix $\Freq{\W}_{pq}\in\mathbb{C}^{\DFTLength\times\DFTLength}$ denotes the \RevFour{\ac{STFT}}-domain representation of \eqref{eq:convolution_matrix}.}

By inserting \eqref{eq:circMat_fromFreqW} into \eqref{eq:windowed_circularMat}, we obtain for a \Walter{submatrix} of the \Walter{time-domain} demixing filter matrix
%\begin{equation}
%\W_{pq}\transp = \DFTWindowDR \FourierMatInv \Freq{\W_{pq}} \FourierMat \DFTWindowRL
%\end{equation}
\Walter{(see Fig. \ref{fig:extend_demixingMatrix2Circular} for an illustration)}
\begin{align}
\W_{pq} &=  (\DFTWindowRL)\transp\FourierMat \Freq{\W}_{pq} \FourierMatInv (\DFTWindowDR)\transp\\[1ex]
&=  \Window_{2L\times \DFTLength}^{1_{2L} 0}\FourierMat \Freq{\W}_{pq} \FourierMatInv \Window_{\RevTwo{\FreqixMax}\times D}^{1_{D}0}. \notag
\end{align}
Now, the output signal $q$ can be written as
\begin{align}
\mathbf{y}_q(\TriBlockIdx) &=\sum_{p=1}^{P}\mathbf{W}_{\RevTwo{pq}}^\text{T}\mathbf{x}_p(\TriBlockIdx) \notag\\
&=\sum_{p=1}^{P}\DFTWindowDR \FourierMatInv \Freq{\W}_{pq} \FourierMat \DFTWindowRL\mathbf{x}_p(\TriBlockIdx)
\label{eq:DFT_demixedSignal}
\end{align}
The \RevFour{\ac{STFT}} representation of the microphone signals, padded with $\DFTLength-2L$ zeros is obtained as 
\begin{equation}
	\Freq{\mathbf{x}}_p(n) = \FourierMat \begin{bmatrix} \mathbf{x}_p\transp(n)& \RevTwo{\mathbf{0}_{1\times(\DFTLength-2L)}}\end{bmatrix}\transp= \FourierMat \DFTWindowRL \mathbf{x}_p.
\end{equation}
\Walter{Note that this step ensures the realization of linear convolution instead of circular convolution.} Identifying the \RevFour{\ac{STFT}} representation of the microphone signals in \eqref{eq:DFT_demixedSignal} yields
\begin{align}
\mathbf{y}_q(\TriBlockIdx) &=\sum_{p=1}^{P}\DFTWindowDR \FourierMatInv \Freq{\W}_{pq} \Freq{\mathbf{x}}_p(\TriBlockIdx) \notag\\
&= \DFTWindowDR\FourierMatInv \begin{bmatrix}\Freq{\W}_{1q},\dots,\Freq{\W}_{Pq}\end{bmatrix}\underbrace{\begin{bmatrix}\Freq{\mathbf{x}}_1(\TriBlockIdx)\\ \vdots \\ \Freq{\mathbf{x}}_P(\TriBlockIdx)\end{bmatrix}}_{\Freq{\mathbf{x}}(\TriBlockIdx)}\\
&= \DFTWindowDR\FourierMatInv \Freq{\W}_{q}\Freq{\mathbf{x}}(\TriBlockIdx). \notag
\end{align}
\Walter{We note that the} multiplication of the window matrices $(\DFTWindowLD)\transp$ and $(\DFTWindowRL)\transp$ yields again a window matrix
\begin{align}
\left(\DFTWindowLD\right)\transp\left(\DFTWindowRL\right)\transp &= \begin{bmatrix} \mathbf{I}_{D}\\ \mathbf{0}_{(2L-D)\times D}
\end{bmatrix}\transp\begin{bmatrix} \mathbf{I}_{2L}\\ \mathbf{0}_{(\DFTLength-2L)\times 2L}
\end{bmatrix}\transp \notag\\[1ex]
&= \DFTWindowDR.
\end{align}
Using this result, we \RevFour{formulate} the \ac{TRINICON} cost function \Walter{\eqref{eq:TRINICON_costFunction}} equivalently in \RevTwo{the} \RevFour{\ac{STFT}} domain \RevFour{by} \eqref{eq:TRINICON_costFct_DFT}. \addtocounter{equation}{1}
%-----------------------------------------------------------------
\subsection{Reformulation of IVA}
%-----------------------------------------------------------------
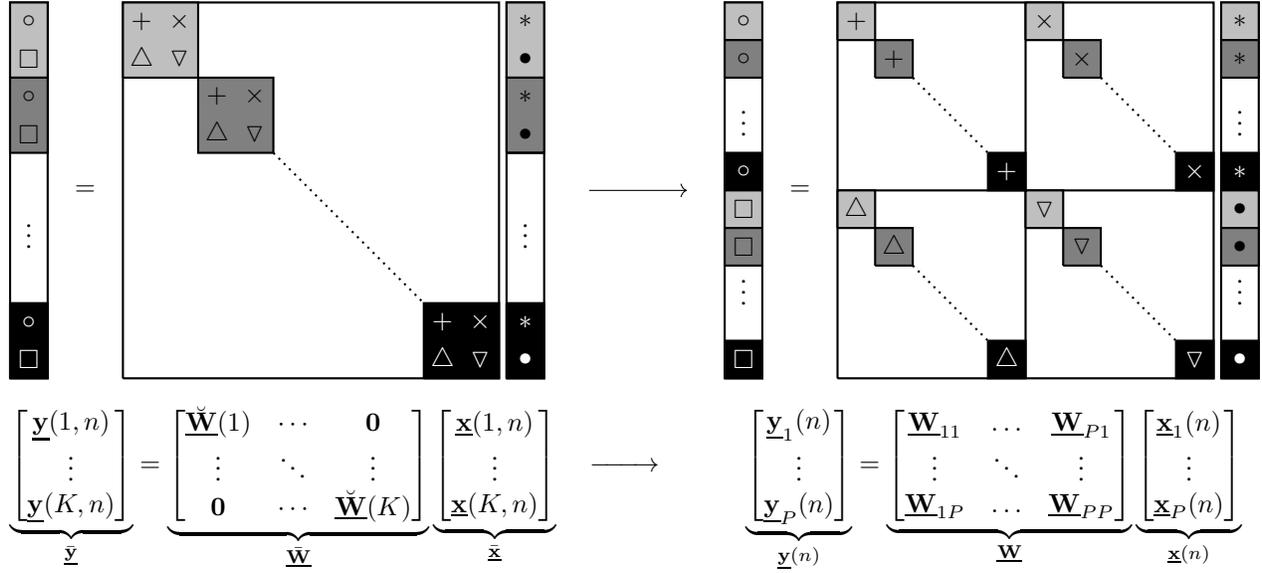
\begin{figure*}
	\centering
	\begin{tikzpicture}
	
%	\tikzstyle{every node}=[draw]
	
	\draw[thick] (0,0)--(0,5)--(0.5,5)--(0.5,0)--(0,0);
	\draw[thick,fill = lightgray] (0,4)--(0.5,4)--(0.5,5)--(0,5)--(0,4);
	\node at (0.25,4.75)(){$\circ$};
	\node at (0.25,4.25)(){$\square$};
	\draw[thick,fill = gray] (0,3)--(0.5,3)--(0.5,4)--(0,4)--(0,3);
	\node at (0.25,3.75)(){$\circ$};
	\node at (0.25,3.25)(){$\square$};
	\node at (0.25,2)(){$\vdots$};
	\draw[thick,fill = black] (0,1)--(0.5,1)--(0.5,0)--(0,0)--(0,1);
	\node at (0.25,0.75)(){\textcolor{white}{$\circ$}};
	\node at (0.25,0.25)(){\textcolor{white}{$\square$}};
	
	\node at (1,2.5)(){$=$};
	
	\draw[thick] (1.5,0)--(1.5,5)--(6.5,5)--(6.5,0)--(1.5,0);
	\draw[thick,fill = lightgray] (1.5,4)--(2.5,4)--(2.5,5)--(1.5,5)--(1.5,4);
	\node at (1.75,4.75)(){$+$};
	\node at (2.25,4.75)(){$\times$};
	\node at (1.75,4.25)(){$\triangle$};
	\node at (2.25,4.25)(){$\triangledown$};
	\draw[thick,fill=gray] (2.5,4)--(3.5,4)--(3.5,3)--(2.5,3)--(2.5,4);
	\node at (2.75,3.75)(){$+$};
	\node at (3.25,3.75)(){$\times$};
	\node at (2.75,3.25)(){$\triangle$};
	\node at (3.25,3.25)(){$\triangledown$};
	\draw[thick,dotted] (3.5,3)--(5.5,1);
	\draw[thick,fill=black] (5.5,0)--(5.5,1)--(6.5,1)--(6.5,0)--(5.5,0);
	\node at (5.75,0.75)(){\textcolor{white}{$+$}};
	\node at (6.25,0.75)(){\textcolor{white}{$\times$}};
	\node at (5.75,0.25)(){\textcolor{white}{$\triangle$}};
	\node at (6.25,0.25)(){\textcolor{white}{$\triangledown$}};
	
	\draw[thick] (6.6,0)--(6.6,5)--(7.1,5)--(7.1,0)--(6.6,0);
	\draw[thick,fill = lightgray] (6.6,4)--(7.1,4)--(7.1,5)--(6.6,5)--(6.6,4);
	\node at (6.85,4.75)(){$\ast$};
	\node at (6.85,4.25)(){$\bullet$};
	\draw[thick,fill = gray] (6.6,3)--(7.1,3)--(7.1,4)--(6.6,4)--(6.6,3);
	\node at (6.85,3.75)(){$\ast$};
	\node at (6.85,3.25)(){$\bullet$};
	\draw[thick] (6.6,3)--(7.1,3);
	\node at (6.85,2)(){$\vdots$};
	\draw[thick,fill = black] (6.6,1)--(7.1,1)--(7.1,0)--(6.6,0)--(6.6,1);
	\node at (6.85,0.75)(){\textcolor{white}{$\ast$}};
	\node at (6.85,0.25)(){\textcolor{white}{$\bullet$}};
	
	\draw[->](7.7,2.5)--(9,2.5);
	
	\begin{scope}[shift = {(9.5,0)}]
	\draw[thick] (0,0)--(0,5)--(0.5,5)--(0.5,0)--(0,0);
	\draw[thick,fill = lightgray] (0,4.5)--(0,5)--(0.5,5)--(0.5,4.5)--(0,4.5);
	\node at (0.25,4.75)(){$\circ$};
	\draw[thick,fill = gray] (0,4)--(0,4.5)--(0.5,4.5)--(0.5,4)--(0,4);
	\node at (0.25,4.25)(){$\circ$};
	\node at (0.25,3.5)(){$\vdots$};
	\draw[thick,fill = black] (0,2.5)--(0,3)--(0.5,3)--(0.5,2.5)--(0,2.5);
	\node at (0.25,2.75)(){\textcolor{white}{$\circ$}};
	\draw[thick,fill=lightgray] (0,2.5)--(0.5,2.5)--(0.5,2)--(0,2)--(0,2.5);
	\node at (0.25,2.25)(){$\square$};
	\draw[thick,fill=gray] (0,2)--(0.5,2)--(0.5,1.5)--(0,1.5)--(0,2);
	\node at (0.25,1.75)(){$\square$};
	\node at (0.25,1.25)(){$\vdots$};
	\draw[thick,fill=black] (0,0.5)--(0.5,0.5)--(0.5,0)--(0,0)--(0,0.5);
	\node at (0.25,0.25)(){\textcolor{white}{$\square$}};
	
	\node at (1,2.5)(){$=$};
	
	\draw[thick] (1.5,0)--(1.5,5)--(6.5,5)--(6.5,0)--(1.5,0);	
	\draw[thick] (4,0)--(4,5);
	\draw[thick] (1.5,2.5)--(6.5,2.5);
	
	\draw[thick, fill= lightgray] (1.5,4.5)--(1.5,5)--(2,5)--(2,4.5)--(1.5,4.5);
	\draw[thick, fill= gray] (2,4)--(2,4.5)--(2.5,4.5)--(2.5,4)--(2,4);
	\draw[thick, fill= black] (3.5,3)--(3.5,2.5)--(4,2.5)--(4,3)--(3.5,3);
	
	\begin{scope}[shift = {(2.5,0)}]
	\draw[thick, fill= lightgray] (1.5,4.5)--(1.5,5)--(2,5)--(2,4.5)--(1.5,4.5);
	\draw[thick, fill= gray] (2,4)--(2,4.5)--(2.5,4.5)--(2.5,4)--(2,4);
	\draw[thick, fill= black] (3.5,3)--(3.5,2.5)--(4,2.5)--(4,3)--(3.5,3);
	\end{scope}
	
	\begin{scope}[shift = {(0,-2.5)}]
	\draw[thick, fill= lightgray] (1.5,4.5)--(1.5,5)--(2,5)--(2,4.5)--(1.5,4.5);
	\draw[thick, fill= gray] (2,4)--(2,4.5)--(2.5,4.5)--(2.5,4)--(2,4);
	\draw[thick, fill= black] (3.5,3)--(3.5,2.5)--(4,2.5)--(4,3)--(3.5,3);
	\end{scope}
	
	\begin{scope}[shift = {(2.5,-2.5)}]
	\draw[thick, fill= lightgray] (1.5,4.5)--(1.5,5)--(2,5)--(2,4.5)--(1.5,4.5);
	\draw[thick, fill= gray] (2,4)--(2,4.5)--(2.5,4.5)--(2.5,4)--(2,4);
	\draw[thick, fill= black] (3.5,3)--(3.5,2.5)--(4,2.5)--(4,3)--(3.5,3);
	\end{scope}
	
	\node at (1.75,4.75)(){$+$};
	\node at (4.25,4.75)(){$\times$};
	\node at (1.75,2.25)(){$\triangle$};
	\node at (4.25,2.25)(){$\triangledown$};
	
	\node at (2.25,4.25)(){$+$};
	\node at (4.75,4.25)(){$\times$};
	\node at (2.25,1.75)(){$\triangle$};
	\node at (4.75,1.75)(){$\triangledown$};
	
	\node at (6.25,2.75)(){\textcolor{white}{$\times$}};
	\node at (3.75,2.75)(){\textcolor{white}{$+$}};
	\node at (6.25,0.25)(){\textcolor{white}{$\triangledown$}};
	\node at (3.75,0.25)(){\textcolor{white}{$\triangle$}};
	
	\draw[thick,dotted] (5,4)--(6,3);
	\draw[thick,dotted] (2.5,4)--(3.5,3);
	\draw[thick,dotted] (5,1.5)--(6,0.5);
	\draw[thick,dotted] (2.5,1.5)--(3.5,0.5);
	
	\begin{scope}[shift = {(6.6,0)}]
		\draw[thick] (0,0)--(0,5)--(0.5,5)--(0.5,0)--(0,0);
		\draw[thick,fill = lightgray] (0,4.5)--(0,5)--(0.5,5)--(0.5,4.5)--(0,4.5);
		\node at (0.25,4.75)(){$\ast$};
		\draw[thick,fill = gray] (0,4)--(0,4.5)--(0.5,4.5)--(0.5,4)--(0,4);
		\node at (0.25,4.25)(){$\ast$};
		\node at (0.25,3.5)(){$\vdots$};
		\draw[thick,fill = black] (0,2.5)--(0,3)--(0.5,3)--(0.5,2.5)--(0,2.5);
		\node at (0.25,2.75)(){\textcolor{white}{$\ast$}};
		\draw[thick,fill=lightgray] (0,2.5)--(0.5,2.5)--(0.5,2)--(0,2)--(0,2.5);
		\node at (0.25,2.25)(){$\bullet$};
		\draw[thick,fill=gray] (0,2)--(0.5,2)--(0.5,1.5)--(0,1.5)--(0,2);
		\node at (0.25,1.75)(){$\bullet$};
		\node at (0.25,1.25)(){$\vdots$};
		\draw[thick,fill=black] (0,0.5)--(0.5,0.5)--(0.5,0)--(0,0)--(0,0.5);
		\node at (0.25,0.25)(){\textcolor{white}{$\bullet$}};
	\end{scope}
	
	\end{scope}
	
\end{tikzpicture}\vspace{3pt}
	\begin{equation*}
	\qquad
	\underbrace{\begin{bmatrix}
		\Freq{\y}(1,\BlockIdx)\\
		\vdots\\
		\Freq{\y}(\FreqixMax,\BlockIdx)\\
		\end{bmatrix}}_{\yPerm}=
	\underbrace{\begin{bmatrix}
		\WINST(1)&\cdots&\mathbf{0}\\
		\vdots&\ddots&\vdots\\
		\mathbf{0}&\cdots&\WINST(\FreqixMax)
		\end{bmatrix}}_{\WPerm}
	\underbrace{\begin{bmatrix}
		\Freq{\x}(1,\BlockIdx)\\
		\vdots\\
		\Freq{\x}(\FreqixMax,\BlockIdx)\\
		\end{bmatrix}}_{\xPerm} 
	\quad\xrightarrow\qquad\qquad\quad
	\underbrace{\begin{bmatrix}
		\Freq{\y}_1(\BlockIdx)\\
		\vdots\\
		\Freq{\y}_P(\BlockIdx)
		\end{bmatrix}}_{\Freq{\y}(\BlockIdx)} = \underbrace{\begin{bmatrix}
		\Freq{\W}_{11}&\dots&\Freq{\W}_{P1}\\
		\vdots&\ddots&\vdots\\
		\Freq{\W}_{1P}&\dots&\Freq{\W}_{PP}
		\end{bmatrix}}_{\WIVA}\underbrace{\begin{bmatrix}
		\Freq{\x}_1(\BlockIdx)\\
		\vdots\\
		\Freq{\x}_P(\BlockIdx)
		\end{bmatrix}}_{\Freq{\x}(\BlockIdx)}\qquad\qquad
	\end{equation*}
	\caption{Illustration of the permutation \RevTwo{in} the \ac{IVA} demixing process. The left rectangle in both pictures represents the demixed signal vector, the square the demixing matrix and the right rectangle the microphone vector for an example of $2$ sources and $2$ channels. Different symbols represent different matrix/vector elements and different gray scales represent different frequencies. \Walter{The corresponding equations \eqref{eq:IVA_demixingComplete} and \eqref{eq:IVA_demixing_permuted} are shown below the corresponding illustration.}}
	\label{fig:resort_IVA_demixing_matrix}
\end{figure*}
In the following section, we will transform the \ac{IVA} cost function into a representation, which allows \RevTwo{a} direct comparison with the \ac{TRINICON} cost function represented in \RevTwo{the} \RevTwo{\ac{STFT}} domain \eqref{eq:TRINICON_costFct_DFT}.

First, we \Walter{observe} that the demixing operation of \ac{IVA} (and also \ac{FD-ICA}) can be described by \RevFour{combining \eqref{eq:ICA_demixing} for all \mbox{$\Freqix\in\{1,\dots,\FreqixMax\}$}} 
%(a more explicit description is \Walter{given} in \eqref{eq:IVA_demixingCompleteBefore})
\begin{equation}
	\underbrace{\begin{bmatrix}
	\Freq{\y}(1,\BlockIdx)\\
	\vdots\\
	\Freq{\y}(\FreqixMax,\BlockIdx)\\
	\end{bmatrix}}_{\yPerm}=
	\underbrace{\begin{bmatrix}
	\WINST(1)&\cdots&\mathbf{0}\\
	\vdots&\ddots&\vdots\\
	\mathbf{0}&\cdots&\WINST(\FreqixMax)
	\end{bmatrix}}_{\WPerm}
    \underbrace{\begin{bmatrix}
	\Freq{\x}(1,\BlockIdx)\\
	\vdots\\
	\Freq{\x}(\FreqixMax,\BlockIdx)\\
	\end{bmatrix}}_{\xPerm}.
\label{eq:IVA_demixingComplete}
\end{equation}
To \Walter{obtain} the same structure of the demixing operation as for the \Walter{\RevFour{\ac{STFT}}-domain} description of \ac{TRINICON}, we multiply \eqref{eq:IVA_demixingComplete} with a permutation matrix $\mathbf{P}$ from the left
\begin{equation}
\mathbf{P}\yPerm=
\mathbf{P}\WPerm\xPerm.
\label{eq:extend_perm_y}
\end{equation}
Now, we use the fact that permutation matrices are nonsingular and $\Perm^{-1} = \Perm\transp$ holds \RevTwo{\cite{harville_matrix_2008}}
%	, and we insert $\mathbf{I} = \Perm\Perm^{-1} = \Perm\Perm\transp$ into \RevTwo{\eqref{eq:extend_perm_y}}
\begin{equation}
\mathbf{P}\yPerm=
\mathbf{P}\WPerm\mathbf{P}^{-1}\mathbf{P}\xPerm.
\end{equation}
\Walter{The equalities} \RevTwo{(see \eqref{eq:determinant_permutationMat} for similar calculations)}
\begin{align}
\left\vert \prod_{\Freqix = 1}^{\FreqixMax}\det \WINST(\Freqix) \right\vert &=\left\vert \det\WPerm \right\vert=\left\vert \det \mathbf{P}\det\WPerm\det\mathbf{P}^{-1} \right\vert \notag\\
&=\left\vert \det \mathbf{P}\WPerm\mathbf{P}^{-1} \right\vert=\left\vert \det \WIVA \right\vert,
\end{align}
\Walter{show} that the \Walter{absolute value of the} determinant of the demixing matrix does not change \Walter{when applying} the permutation matrices. The \Walter{rearrangement of the submatrices by the permutation matrices} is illustrated in Fig.~\ref{fig:resort_IVA_demixing_matrix}.

The vector of the $q$th channel containing all \RevTwo{$\FreqixMax$} frequency \RevTwo{bins} is defined as
\begin{equation}
\Freq{\x}_q(\BlockIdx) = \left[\Freq{x}_q(1,\BlockIdx),\dots,\Freq{x}_q(\FreqixMax,\BlockIdx)\right]\transp
\end{equation}
and the corresponding vector of demixed signals as
\begin{equation}
\Freq{\y}_q(\BlockIdx) = \left[\Freq{y}_q(1,\BlockIdx),\dots,\Freq{y}_q(\FreqixMax,\BlockIdx)\right]\transp.
\end{equation}
With these definitions, the demixing process at time frame $\BlockIdx$ can be written in matrix notation as
\begin{equation}
\underbrace{\begin{bmatrix}
	\Freq{\y}_1(\BlockIdx)\\
	\vdots\\
	\Freq{\y}_P(\BlockIdx)
	\end{bmatrix}}_{\Freq{\y}(\BlockIdx)} = \underbrace{\begin{bmatrix}
	\Freq{\W}_{11}&\dots&\Freq{\W}_{P1}\\
	\vdots&\ddots&\vdots\\
	\Freq{\W}_{1P}&\dots&\Freq{\W}_{PP}
	\end{bmatrix}}_{\WIVA}\underbrace{\begin{bmatrix}
	\Freq{\x}_1(\BlockIdx)\\
	\vdots\\
	\Freq{\x}_P(\BlockIdx)
	\end{bmatrix}}_{\Freq{\x}(\BlockIdx)}
\label{eq:IVA_demixing_permuted}
\end{equation}
with subblock $\Freq{\W}_{pq}$ containing all frequency \Walter{bins} defined as 
\begin{equation}
\Freq{\W}_{pq}=\begin{bmatrix}
\Freq{w}_{pq}(1)&&\mathbf{0}\\
&\ddots&\\
\mathbf{0}&&\Freq{w}_{pq}(\FreqixMax)
\end{bmatrix}.
\end{equation}
The transformation of the \ac{PDF} of the input signal vector $\Freq{\x}$ to the output signal vector $\Freq{\y}$ is again accomplished by a change of variables
\begin{equation}
p(\Freq{\y}(\BlockIdx)) = \frac{1}{\left\vert \det \WIVA\right\vert^2}\ p(\Freq{\x}(\BlockIdx)).
\end{equation}

By applying the same rules for a \Walter{linear mapping} of \RevTwo{complex-valued random} variables, we can show that the \ac{PDF} of the demixed signals is not affected by the permutation
\begin{align}
	p(\yPerm(\BlockIdx)) &=p\left(\mathbf{P}\left[\Freq{\y}\transp(1,\BlockIdx),\dots,\Freq{\y}\transp(\FreqixMax,\BlockIdx)\right]\transp\right)\\[1ex]
	&=\frac{1}{\left\vert \det \mathbf{P} \right\vert^2}\ p\left(\y(\BlockIdx)\right)=p\left(\y(\BlockIdx)\right). \notag
\end{align}
The same holds for $p\left(\Freq{\x}\right)$. Hence, we can \RevTwo{write} the \ac{IVA} cost function \Walter{representing} the mutual information of the output channels equivalently as
\begin{align}
&\RevThree{\tilde{J}_\text{IVA}}(\WIVA) = \mathcal{KL} \left\lbrace p(\Freq{\y}(\BlockIdx)) \bigg \vert \bigg \vert  \prod_{q=1}^{P} p(\Freq{\y}_q(\BlockIdx))\right\rbrace\\
&=  \underbrace{\mathcal{H}_{p(\Freq{\x}(\BlockIdx))}}_\text{const.} -\sum_{q=1}^{P} \mathcal{E}_{\RevFour{\text{frame}}}\left[\log p(\Freq{\y}_q(\BlockIdx) )\right] - 2\log \vert  \det\WIVA \vert . \notag
\end{align}
After neglecting constant terms and approximating the expectation operator \Walter{by averaging \RevThree{over a time interval of $\NumBlocks$ time-frequency bins along the time axis for each channel $q$}}, we arrive at
\begin{equation}
\RevTwo{J_\text{IVA}(\WIVA) \RevThree{=} \sum_{q=1}^{P}\RevThree{\frac{1}{\NumBlocks}\sum_{\BlockIdx=0}^{\NumBlocks-1}}\left\lbrace  -\log p(\Freq{\y}_q(\BlockIdx))\right\rbrace+2\log\left\vert \det \WIVA \right\vert,}
\end{equation}
or more explicitly \RevTwo{at}
\begin{align}
	J_\text{IVA}(\WIVA) &\RevTwo{\RevThree{=}  \sum_{q=1}^{P}\RevThree{\frac{1}{\NumBlocks}\sum_{\BlockIdx=0}^{\NumBlocks-1}}\left\lbrace- \log p(\Freq{\W}_{q}\Freq{\mathbf{x}}(\BlockIdx))\right\rbrace \dots \notag}\\[1ex]
	&\qquad+2\log\left\vert \det \begin{bmatrix}
		\Freq{\W}_{11}&\dots&\Freq{\W}_{P1}\\
		\vdots&\ddots&\vdots\\
		\Freq{\W}_{1P}&\dots&\Freq{\W}_{PP}
	\end{bmatrix} \right\vert,
	\label{eq:costFunction_IVAreformExplicit}
\end{align}
where \RevTwo{$\Freq{\W}_q$ represents the $q$th row of $\Freq{\W}$.} \RevThree{Note that \eqref{eq:costFunction_IVA_approx} and \eqref{eq:costFunction_IVAreformExplicit} are identical up to the representation of the demixing matrix.}
%\begin{align}
%\Freq{\mathbf{y}}_q(\BlockIdx) &=\sum_{p=1}^{P} \Freq{\W_{pq}} \Freq{\mathbf{x}}_p(\BlockIdx)= \underbrace{\begin{bmatrix}\Freq{\W}_{1q},\dots,\Freq{\W}_{Pq}\end{bmatrix}}_{\Freq{\W}_{q}}\underbrace{\begin{bmatrix}\Freq{\mathbf{x}}_1(n)\\ \vdots \\ \Freq{\mathbf{x}}_P(n)\end{bmatrix}}_{\Freq{\mathbf{x}}(\BlockIdx)}\notag\\
%&= \Freq{\W}_{q}\Freq{\mathbf{x}}(\BlockIdx).
%\end{align}
%-----------------------------------------------------------------
\subsection{Relation between \ac{IVA} and \ac{TRINICON}}
\label{sec:trinicon_iva}
%-----------------------------------------------------------------
In \RevTwo{the following}, we \RevTwo{investigate the} relations between the \ac{IVA} cost function \eqref{eq:costFunction_IVAreformExplicit} and the \ac{TRINICON} cost function \eqref{eq:TRINICON_costFct_DFT}.
% \RevTwo{Please note, that the \ac{TRINICON} cost function is written in dependence of sample indices and the \ac{IVA} cost function in dependence of \ac{STFT} frame indices $n$. The relation between both is defined by the mapping \eqref{eq:mapping_BlockFrame}.} 
First of all, \RevThree{in addition to the} summation over \RevFour{multiple frames} \RevTwo{$\TriBlockIdx$}, \RevFour{we observe} in the \ac{TRINICON} cost function \eqref{eq:TRINICON_costFct_DFT} \RevFour{an expectation over the log-likelihood of output samples in the time domain, which calls for approximation by a sample-wise averaging}. This \Walter{represents} the exploitation of nonstationarity \Walter{of the observations} in \ac{TRINICON} \RevThree{as statistics at different points in time are considered}. To simplify the comparison, we eliminate this \RevThree{expectation} in the following, i.e., we \RevThree{use instantaneous realizations of the demixed signal vectors}. By inspection of \RevTwo{\eqref{eq:TRINICON_costFct_DFT} and \eqref{eq:costFunction_IVAreformExplicit}}, we can \RevTwo{furthermore} identify two main components \Walter{in both \RevFour{cost functions}}: \RevFour{a term reflecting a source model} and a log-det term including \Walter{only} the demixing matrix. \Walter{We compare these two terms \RevTwo{individually}.}

\Walter{Starting} with the source model \Walter{term, we obtain for} \ac{TRINICON} \Walter{under the \RevThree{sums over $\TriBlockIdx$ and} $q$}
\begin{equation}
	\log p\left(\DFTWindowDR\FourierMatInv\Freq{\W}_{q}\Freq{\mathbf{x}}(\TriBlockIdx)\right)
\end{equation}
and for \ac{IVA}
\begin{equation}
	\log p(\Freq{\W}_{q}\Freq{\mathbf{x}}(\BlockIdx)).
\end{equation}
By neglecting the window matrix $\DFTWindowDR$, i.e., by using a circular convolutive mixture model for \ac{TRINICON}, both terms are identical up to \Walter{a} \ac{DFT}. The \ac{DFT} matrix is a unitary matrix\Walter{, i.e., $\vert \det \FourierMat\vert = 1$}, hence
\begin{gather}
	p\left(\FourierMatInv\Freq{\W}_{q}\Freq{\mathbf{x}}(\TriBlockIdx)\right) = \vert \det \FourierMat\vert^2 p\left(\Freq{\W}_{q}\Freq{\mathbf{x}}(\TriBlockIdx)\right)= p\left(\Freq{\W}_{q}\Freq{\mathbf{x}}(\TriBlockIdx)\right).\raisetag{-2pt}
\end{gather}
%\Walter{By using the mapping $\BlockIdx = f(\TriBlockIdx)$ \RevTwo{between the sample index $m$ and the \ac{STFT} time frame $\BlockIdx$ defined in \eqref{eq:mapping_BlockFrame}}, we see that} 
\RevTwo{We see that} the term corresponding to the source model is \RevFour{equivalent} for \ac{TRINICON} and \ac{IVA} up to the windowing ensuring the linear \Walter{convolution} model for \Walter{demixing by} \ac{TRINICON}.

\RevThree{The} log-det term reads for \ac{TRINICON} after removing the window matrices ensuring linear convolution
\begin{align}
	&\log\left\vert \det \left(\mathbf{I}_P \otimes\FourierMat \right)  \Freq{\W} \left(\mathbf{I}_P \otimes \FourierMatInv\right) \right\vert = \notag\\[0.5ex]
	&=\log\left\vert \det \left(\mathbf{I}_P \otimes\FourierMat \right)  \det\Freq{\W} \det\left(\mathbf{I}_P \otimes \FourierMatInv\right) \right\vert = \notag\\[0.5ex]
	&=\log\left\vert \det(\mathbf{I}_P)^P (\det\FourierMat)^\FreqixMax  \det\Freq{\W} \det(\mathbf{I}_P)^P \det( \FourierMatInv)^\FreqixMax \right\vert= \notag\\[0.5ex]
	&=\log\left\vert \det\Freq{\W} \right\vert.
	\label{eq:rephrased_logDet}
\end{align}
Hereby, we used \RevTwo{the property} that for an $a\times a$ matrix $\mathbf{A}$ and a $b\times b$ matrix $\mathbf{B}$ the following holds $\det\left(\mathbf{A}\otimes\mathbf{B}\right) = \left(\det\mathbf{A}\right)^a\left(\det\mathbf{B}\right)^b$.
The last term \RevThree{in \eqref{eq:rephrased_logDet} corresponds to the} log-det term of the \ac{IVA} cost function, \Walter{except \RevTwo{for} the factor $2$. Note that the \ac{IVA} cost function for real-valued signals does not contain this factor $2$ \cite{anderson_independent_2014} and \RevTwo{then} exact equality \RevThree{is given}.}

\RevThree{To summarize,} the window matrices \RevThree{ensuring} a linear \Walter{convolution for the} demixing system for \ac{TRINICON} \RevThree{\RevFour{constitute the} main difference between \ac{TRINICON} and \ac{IVA}. These \RevFour{window matrices}} disappear \Walter{for} the circular \Walter{convolution for the} demixing system of \ac{IVA} (see Sec.~\ref{sec:relation_BSSAlgorithms} for a discussion of its implications). \RevThree{\RevFour{If \RevFive{the} window matrices are dropped}, the log-det term is essentially the same (up to a factor of $2$ stemming from the formulation of \ac{IVA} in the frequency domain) for \ac{IVA} and \ac{TRINICON}. The source terms are also the same when the additional expectation in \ac{TRINICON} is specialized to instantaneous realizations. \RevFour{\RevTwo{\ac{IVA} can therefore be viewed as a special case of \ac{TRINICON} \RevThree{\RevFive{resulting from} omitting the extra expectation and \RevFour{reducing the linear convolution to a circular one by} dropping} $\DFTWindowDR$.}}

For \ac{IVA}, the source model is often specialized to uncorrelated (but not statistically independent frequency bins) \cite{kim_blind_2007}\RevFour{, which allows to optimize the demixing filter frequency bin-wise}. \RevFour{For \ac{TRINICON} typically correlation between successive signal samples is modeled and, hence, broadband demixing filters are optimized.}}
%By choosing a Gaussian source model, \ac{SOS} \ac{TRINICON} (see Sec.~\ref{sec:trinicon_sosMethods}) is obtained yielding, which is shown to work robustly in practice.} 
%-----------------------------------------------------------------
\subsection{Relation Between \ac{TRINICON} and \ac{SOS}-based Methods}
\label{sec:trinicon_sosMethods}
%-----------------------------------------------------------------
In the previous section, we showed \RevTwo{that the \ac{IVA} cost function is a special case of the \ac{TRINICON} cost function} \RevFour{up to an additional factor of two originating from the representation of complex-valued random numbers. To this end, we simplified the \ac{TRINICON} model by} relaxation of the linear demixing model to a circulant one and by dropping the exploitation \RevFour{of nonstationarity by an additional expectation over time-domain output samples}. In this section, we want to investigate the form of the cost function if the exploitation of nongaussianity is dropped, i.e., only \ac{SOS} are considered. To this end, we rewrite the \ac{TRINICON} cost function \eqref{eq:trinicon_costFunctionKL} such that it is expressed as a sum of differential entropies of output signal blocks \RevTwo{\cite{cover_elements_2006}}
\begin{align}
&\RevFour{\tilde{J}}_{\text{TRI}}(\mathbf{W}) = \frac{1}{\NumBlocks}\sum_{\TriBlockIdx=0}^{\NumBlocks-1}\mathcal{KL}\left\lbrace p(\mathbf{y}(\TriBlockIdx)) \bigg\Vert\prod_{q=1}^{P}p(\mathbf{y}_q(\TriBlockIdx)) \right\rbrace\notag\\
%&\quad= \frac{1}{\NumBlocks}\sum_{\TriniconBlock=0}^{\NumBlocks-1}\int p(\mathbf{y}(\TriniconBlock L)) \log \frac{p(\mathbf{y}(\TriniconBlock L))}{\prod_{q=1}^{P}p(\mathbf{y}_q(\TriniconBlock L))}\mathrm{d}\mathbf{y}(\TriniconBlock L)\notag \\
&\RevTwo{\quad= \frac{1}{\NumBlocks}\sum_{\TriBlockIdx=0}^{\NumBlocks-1}            \left(\sum_{q=1}^{P}\mathcal{H}_{p(\mathbf{y}_q(\TriBlockIdx))}-\mathcal{H}_{p(\mathbf{y}(\TriBlockIdx))}\right).}
\label{eq:trinicon_alternativeCostFct}
\end{align}
To express the exclusive dependency on \ac{SOS}, we choose the output signal vectors to follow a $PD$-variate Gaussian distribution
\begin{equation*}
	\RevTwo{\mathbf{y}(\TriBlockIdx)\sim\mathcal{N}\left(\mathbf{0},\mathbf{R}(\TriBlockIdx)\right).}
\end{equation*}
Due to the statistical independence of the source signals, the covariance \RevTwo{matrix of the output signals should be a blockdiagonal matrix, i.e.,}
\begin{equation}
	\RevTwo{\mathbf{R}(\TriBlockIdx)\ \RevTwo{\overset{!}{=}}\ \text{Bdiag}\{\mathbf{R}_1(\TriBlockIdx),\dots,\mathbf{R}_P(\TriBlockIdx)\},}
\end{equation}
where $\text{Bdiag}\{\cdot\}$ is an operator that generates a block diagonal matrix from its arguments. The $q$-th output signal block is $D$-variate \RevTwo{normally} distributed
\begin{equation}
	\RevTwo{\mathbf{y}_q(\TriBlockIdx)\sim\mathcal{N}\left(\mathbf{0},\mathbf{R}_q(\TriBlockIdx)\right).}
\end{equation}
The differential entropies of these Gaussian distributions are given as \cite{cover_elements_2006}
\begin{equation}
	\RevTwo{\mathcal{H}_{p(\mathbf{y}_q(\TriBlockIdx))} = \frac{1}{2}\log\left[(2\pi e)^D\det\mathbf{R}_q(\TriBlockIdx)\right]}
\end{equation}
\begin{equation}
\RevTwo{	\mathcal{H}_{p(\mathbf{y}(\TriBlockIdx))}= \frac{1}{2}\log\left[(2\pi e)^{PD}\det\mathbf{R}(\TriBlockIdx)\right].}
\end{equation}
\RevTwo{Insertion into} \eqref{eq:trinicon_alternativeCostFct}, yields the cost function for \ac{SOS} \ac{TRINICON}
\begin{align}
	\RevFour{\tilde{J}}_{\text{SOS}}(\mathbf{W})&= \frac{PD}{2}\log(2\pi e)\frac{1}{\NumBlocks}\sum_{\RevTwo{\TriBlockIdx}=0}^{\NumBlocks-1}\dots\\
	&\dots\sum_{q=1}^{P}\log\det\mathbf{R}_q(\RevTwo{\TriBlockIdx})-\log\det\mathbf{R}(\RevTwo{\TriBlockIdx})\notag.
\end{align}
We introduce the $\mathrm{bdiag}\{\cdot\}$ operator, which sets all \RevFour{block-off-diagonals} of the matrix-valued argument to zero, and obtain the following expression, which is proportional to the \ac{SOS} cost function
\begin{align}
	\RevFour{\tilde{J}}_{\text{SOS}}(\mathbf{W})&\propto \sum_{\RevTwo{\TriBlockIdx}=0}^{\NumBlocks-1}\bigg\lbrace\log\det\mathrm{bdiag}\{\mathbf{R}(\RevTwo{\TriBlockIdx})\}-\log\det\mathbf{R}(\RevTwo{\TriBlockIdx})\bigg\rbrace.
	\label{eq:tri_SOS_cost_function}
\end{align}
This result \RevTwo{represents an approximate joint diagonalization problem \cite{pham_joint_2001} and} coincides with the \ac{SOS}-based cost function in \cite{buchner_generalization_2005}. \RevThree{It also} represents a generalization of a well-known cost function for \ac{SOS}-based \ac{BSS} \RevFive{\cite[Theorem 1]{matsuoka_neural_1995}}\RevThree{, where} the nonstationarity of the (speech) \RevTwo{source signals is exploited by considering correlation over multiple time lags by calculating $D\times D$ correlation matrices (i.e., modeling nonwhiteness) at different time instants $\RevTwo{\TriBlockIdx}$.}
\RevFour{While \cite{matsuoka_neural_1995} aims at separating instantaneous signal mixtures, \eqref{eq:tri_SOS_cost_function} deals with convolutive mixtures. However, similar to \ac{FD-ICA}, convolutive \ac{BSS} can be addressed by applying instantaneous \ac{BSS} frequency bin-wise and solving the inner permutation problem. Following this strategy, \cite{matsuoka_neural_1995} has been extended to convolutive mixtures in \cite{murata_approach_2001}. Similar approaches relying on \ac{SOS} but employing slightly different cost functions for approximate joint diagonalization in individual frequency bins have been proposed \cite{parra_convolutive_2000,ikram_permutation_2005,chabriel_joint_2014}.}
\subsection{Summary - Relation Between Convolutive \ac{BSS} Algorithms}
\label{sec:relation_BSSAlgorithms}
\begin{figure}
	\begin{center}
		\begin{tikzpicture}[node distance=1cm, auto,]
		%nodes
		\node[punkt] (trinicon) {\ac{TRINICON}};
		\node[below=2.2cm of trinicon](SOStriniconIVA) {};
		\node[punkt, inner sep=5pt,left=1.075cm of SOStriniconIVA](SOStrinicon) {\ac{SOS} \ac{TRINICON}};
		\node[punkt, inner sep=5pt,right=0.2cm of SOStriniconIVA](iva) {\ac{IVA}};
		\node[punkt, inner sep=5pt,below=2cm of iva](fdica) {\ac{FD-ICA}};
%		\node[punkt, inner sep=5pt,below=1.75cm of fdica](ica) {\ac{ICA}};
		% We make a dummy figure to make everything look nice.
		\node[below=0.75cm of trinicon] (trinicon_rel) {};
		\node[box, right=-0.9cm of trinicon_rel, align=left] (t) {$\bullet$ convolution: linear $\rightarrow$ circular\\ $\bullet$ skip nonstationarity}
		edge[pil,<-,bend right=0] (trinicon.east)
		edge[pil, bend left=0] (iva.north);
		\node[below=0.9cm of iva] (iva_rel) {};
		\node[box, right=-2.08cm of iva_rel, align=left] (t) {source \ac{PDF}:\\ multivariate $\rightarrow$ univariate}
		edge[pil,<-,bend right=0] (iva.south)
		edge[pil, bend left=0] (fdica.north);
%		\node[below=0.75cm of fdica] (fdica_rel) {};
%		\node[box, right=-3cm of fdica_rel, align=left] (t) {multiple bin-wise \ac{ICA} $\rightarrow$ single \ac{ICA}}
%		edge[pil,<-,bend right=0] (fdica.south)
%		edge[pil, bend left=0] (ica.north);
		\node[above=0.75cm of trinicon] (SOStrinicon_rel) {};
		\node[box, left=0.75cm of trinicon_rel, align=left] (t) {Gaussian source \ac{PDF}} edge[pil,<-,bend left=0] (trinicon.west) edge[pil, bend right=0] (SOStrinicon.north);
		\end{tikzpicture}
	\end{center}
	\caption{Illustration of the relations between the discussed \ac{BSS} algorithms.}
	\label{fig:flow_diagram}
\end{figure}
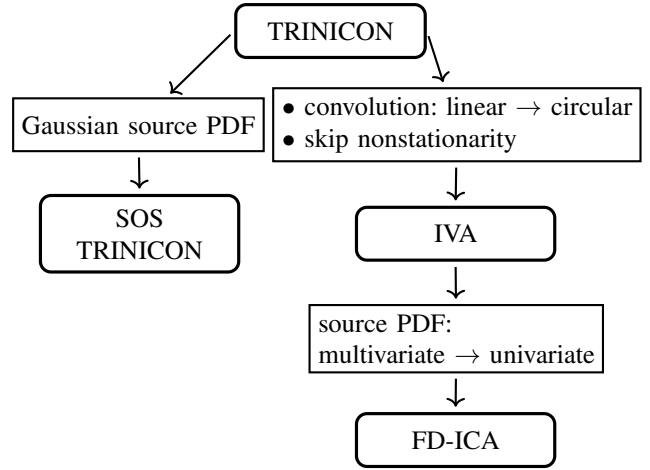

%-----------------------------------------------------------------
\RevFour{The relations between the discussed algorithms for convolutive \ac{BSS} are illustrated in Fig.~\ref{fig:flow_diagram}.} We showed in Section~\ref{sec:trinicon_iva} that (real-valued) \ac{IVA} can be considered as a special case of the \ac{TRINICON} \Walter{BSS} framework. In \RevFour{Section~\ref{sec:IVA}}, we \Walter{summarized} well-known relations from literature, which state that \ac{ICA} is a special case of \ac{FD-ICA}, which is again a special case of \ac{IVA}. Furthermore, we showed the relation of \ac{TRINICON} to \RevThree{other} \ac{SOS}-based \ac{BSS} methods. Hence, \ac{TRINICON} \Walter{is shown to be} the most general approach of the discussed \ac{ICA}-based convolutive \ac{BSS} algorithms.
%\begin{figure}
%	\begin{center}
%		\begin{center}
%			\def\ellipseTrinicon{(0,0) ellipse (10em and 8em)}
%			\def\ellipseIVA{(0.5,0) ellipse (8em and 6em)}
%			\def\ellipseFDICA{(1,0) ellipse (6em and 4em)}
%			\def\ellipseICA{(1.5,0) ellipse (4em and 2em)}
%			\begin{tikzpicture}
%			\draw \ellipseTrinicon;
%			\draw \ellipseIVA;
%			\draw \ellipseFDICA;
%			\draw \ellipseICA;
%			
%			% first coordinate control x axis, second controls y axis
%			\node at (1.5,0) (A) {ICA};
%			\node at (0.9,.9) (B) {FD-ICA};
%			\node at (-0.2,1.5) (C) {IVA};
%			\node at (-1,2.2) (D) {TRINICON};
%			
%			\end{tikzpicture}
%		\end{center}
%	\end{center}
%	\caption{Illustration of the relations between the discussed \ac{BSS} algorithms.}
%	\label{fig:venn_diagram}
%\end{figure}
%%%%%%%%%%%%%%%%%%%%%%%%%%%%%%%%%%%%%%%%%%%%%%%%%%%%%%%%%%%%%%%%%%%%%%%%%%%%
\section{\RevTwo{Impact} of Model Differences}
\label{sec:pros_cons}
%%%%%%%%%%%%%%%%%%%%%%%%%%%%%%%%%%%%%%%%%%%%%%%%%%%%%%%%%%%%%%%%%%%%%%%%%%%%
In this section, we \RevFour{discuss the impact of} the identified differences between the algorithms \RevTwo{and quantify their influence by experiments.}
%-----------------------------------------------------------------
\subsection{Exploited Signal Properties \RevFour{and Source Models}}
\label{sec:signal_properties}

For the typical source model applied \RevTwo{if} \ac{IVA} \RevTwo{is used for audio mixtures}, the frequency bins are modeled to follow a supergaussian \ac{PDF} and to be statistically dependent but uncorrelated \cite{kim_blind_2007}, i.e., \RevFour{weak-sense} nonwhiteness \RevFour{across} frequency bins is not \RevTwo{taken into account}. In \ac{FD-ICA}, the frequency bins are treated independently, i.e., \RevFour{the frequency bins are assumed to be strictly white and, hence,} nonwhiteness is not \RevTwo{modeled} here \RevTwo{either}. \RevTwo{However, even if \RevThree{\RevFour{weak-sense} whiteness is typically assumed} for \ac{IVA} applied to acoustic mixtures \RevThree{\cite{kim_blind_2007,ono_stable_2011}}, the assumption of \RevFour{weak-sense} whiteness of the demixed signal vectors is not necessary (see, e.g., \cite{anderson_joint_2012})\RevFour{, i.e., the demixed signals may be correlated across frequency bins}.}

\RevFour{Desired acoustic signals are often spectrally structured, i.e., nonwhite, which is ignored by the above-mentioned source models for \ac{IVA}. Hence, powerful extensions of \ac{IVA} incorporate models that explicitly model the spectral structure of the source signals. A widely used extension is \ac{ILRMA} \cite{kitamura_determined_2016} which leverages \ac{NMF} \cite{fevotte_nonnegative_2009} for modeling frequency-variant signal variances. The expressive capabilities of deep learning has been exploited for building pre-trained source models for \ac{IVA}, e.g., \cite{makishima_independent_2019}.}

%In both cases, the frequency bin-wise or frame-wise signals are assumed to be i.i.d. over time, which implies the assumption of temporally white and stationary signals. However, practical implementations usually use overlapping frames, which may represent, dependent on the overlap, nonwhiteness up to a certain \RevTwo{extent}. A very large overlap could promote the exploitation of nonwhiteness and nonstationarity. This would yield \ac{BSS} algorithms based on \ac{IVA} or \ac{FD-ICA}, which use nonwhiteness and nonstationarity in addition to nongaussianity. However, this is part of further investigations.
%
%\RevTwo{
%	[add second summation]
%	\begin{align}
%	&J_\text{IVA}(\WINST(\Freqix)) \overset{c}{=}\\
%	&\quad -\frac{1}{\NumBlocks}\sum_{\BlockIdx=0}^{\NumBlocks-1}\left\lbrace \sum_{q=1}^{P}\log p(\Freq{\y}_q(\BlockIdx))+2\sum_{\Freqix=1}^{\FreqixMax}\log\left\vert \det \WINST(\Freqix) \right\vert\right\rbrace\notag
%\end{align}}
\ac{TRINICON} is capable to estimate separated sources \RevTwo{exploiting multiple} source properties and separation is possible even if not all signal properties are fulfilled \RevTwo{\cite{aichner_acoustic_2007}}. \RevTwo{It} has been shown that modeling of all three signal properties \RevThree{yields} \RevTwo{faster convergence speed} \cite{buchner_blind_2003} \RevTwo{relative to \ac{TRINICON} variants taking fewer properties into account}. \RevFive{However, for \ac{TRINICON}, nonstationarity can be considered as the most important signal property in terms of a criterion for separation (see \cite{brendel_diss} for a detailed discussion).} \ac{FD-ICA} and \ac{IVA} on the other hand, \RevTwo{usually} rely on nongaussianity \RevTwo{if applied to acoustic mixtures and are likely to} fail if this signal property is not fulfilled \RevTwo{\cite{hyvarinen_independent_2001}}.
%-----------------------------------------------------------------
\subsection{Ambiguities}
%-----------------------------------------------------------------
The inner permutation problem, i.e., the permutation of the ordering of the output channels \RevTwo{among frequency bins}, is a well-known problem of\RevTwo{, e.g.,} \ac{FD-ICA} \cite{vincent_audio_2018}\RevTwo{,} and has been addressed by numerous repair mechanisms \RevThree{applied as postprocessing step to the separated frequency bins} \cite{sawada_robust_2004,vincent_audio_2018}. \RevTwo{In contrast, \ac{IVA} aims at solving this issue \RevThree{directly} on cost function level by introducing} a multivariate source \ac{PDF} \cite{kim_blind_2007}. However, for auxIVA \cite{ono_stable_2011}, \RevTwo{block permutation} effects have been observed \cite{liang_overcoming_2012}, i.e., the inner permutation problem is only solved for frequency bins which are close to each other. For \RevThree{distant} frequency bins, the statistical coupling is decreased \cite{liang_overcoming_2012} which causes the permutation of frequency blocks. For \ac{TRINICON}, no inner permutation problems \RevFive{occur} as the cost function is defined in time domain, i.e., based on a broadband signal model.

%The \RevTwo{outer} permutation ambiguity of the \RevTwo{broadband} output signals occurs for all \ac{BSS} algorithms. To solve this issue, additional information about the scenario \RevTwo{is required}, e.g., the direction of arrival of the sources \cite{vincent_geometrically_2015,brendel_spatially_2019,reindl_minimum_2014,brendel_journal} or reference signals statistically \RevTwo{linked to} the source signals \cite{nesta_supervised_2017}.

Finally, the so-called filtering ambiguity occurs, i.e., the source signals can only be estimated up to \RevTwo{an} arbitrary filtering \RevTwo{\cite{aichner_acoustic_2007}}. This is reflected by the fact that \ac{TRINICON} \ac{BSS} is \RevFive{estimating the source signals in a filtered (e.g., reverberated version)} and does not \RevTwo{necessarily} \RevThree{equalize (dereverberate)} (see \cite{buchner_trinicon_2010} for the extension to joint dereverberation and separation). For \ac{FD-ICA} and \ac{IVA}, the filtering ambiguity appears as an arbitrary complex (including magnitude and phase) scaling in each frequency bin \cite{vincent_audio_2018}, which has to be \RevTwo{addressed} prior to transforming the separated signals back to time \RevTwo{domain,} e.g., by the minimum distortion principle \cite{matsuoka_minimal_2002}. \RevFour{\RevFive{For implementations of \ac{TRINICON}, the} filtering ambiguity is often addressed by fixing the
diagonal filters to represent a pure delay, i.e., by estimating \acp{RTF} with the remaining filters \RevFive{\cite{meier_analysis_2015}}.}

%As \ac{TRINICON} \RevTwo{estimates} broadband filters, \RevTwo{repair mechanisms to solve a bin-wise scaling problem} are not needed.
%-----------------------------------------------------------------
\subsection{Optimization Strategies}
%-----------------------------------------------------------------
To iteratively minimize the \ac{FD-ICA} cost function \eqref{eq:costFunction_FDICA_approx}, usually a gradient descent algorithm is applied \cite{bell_information-maximization_1995}. The gradient update is very often replaced by the natural gradient \cite{amari_natural_1998}, which \RevFive{generally yields} faster convergence due to the fact that the geometric structure of the space of possible solutions is taken into account. Besides gradient-based optimization approaches, also Newton-type algorithms \cite{palmer_newton_2008} or \ac{MM}-type algorithms \cite{vigneron_auxiliary-function-based_2010} are applied for optimization. A well-known optimization strategy for \ac{FD-ICA} is the FastICA algorithm \cite{hyvarinen_independent_2001} \RevTwo{which can be interpreted as a Newton-type optimization method} \RevThree{\cite{hyvarinen_fast_1999}}.

\RevTwo{For minimizing the \ac{IVA} cost function \eqref{eq:costFunction_IVA_approx}, similar techniques as for \ac{FD-ICA} have been proposed, e.g.,} natural gradient \cite{kim_blind_2007}. \RevFour{In particular, \ac{MM}-based optimization algorithms \cite{ono_stable_2011,scheibler_fast_2020,scheibler_independent_2021,brendel_accelerating_nodate,brendel_faster_nodate} are used for \ac{IVA} to separate acoustic sources due to their robust and fast convergence.} In addition to that, a fast fixed-point algorithm for \ac{IVA} has been proposed \cite{lee_fast_2007}. \RevFour{The computational complexity of various state-of-the-art optimization approaches for \ac{IVA} is analyzed in, e.g., \cite{scheibler_fast_2020,scheibler_independent_2021}.}

The \ac{TRINICON} cost function is usually \RevTwo{optimized} by a steepest descent technique based on the nonholonomic natural gradient \cite{amari_nonholonomic_2000,aichner_acoustic_2007}. Also a Newton-type algorithm for the optimization of the \ac{TRINICON} cost function has been proposed~\cite{buchner_broadband_2010}.

Frequency-domain approaches are known for their lower computational complexity. Hence, the computational load of \ac{FD-ICA} and \ac{IVA} can be considered to be lower than a general time-domain implementation of \ac{TRINICON}. However, several equivalent frequency-domain realizations of \ac{TRINICON} \cite{aichner_acoustic_2007,buchner_generalization_2005} and real-time strategies based on approximations of the gradient updates \cite{aichner_real-time_2006,anderson_gpu-accelerated_2014} have been proposed to compensate for this. \RevFour{The computational complexity of a time-domain real-time implementation of \ac{TRINICON} is investigated in detail in terms of arithmetic operations in \cite{aichner_real-time_2006}.}

%To summarize, a variety of optimization strategies has been proposed for both \ac{FD-ICA} and \ac{IVA} providing different benefits \RevTwo{and} drawbacks. \RevTwo{On the other hand, the} optimization of \ac{TRINICON} is \RevTwo{so far mostly} based on a natural gradient optimization scheme. Therefore, \ac{TRINICON} could potentially benefit from taking different optimization strategies into account.
%-----------------------------------------------------------------
\subsection{Convolution Model}
%-----------------------------------------------------------------
\RevTwo{Using frequency bin-wise instantaneous} \ac{BSS}, as in \ac{FD-ICA} and \ac{IVA}, and transforming the results back into the time domain by inverse \ac{STFT} corresponds to a circular convolution in the time domain. \RevThree{In contrast}, \RevTwo{the generic} \ac{TRINICON} \RevTwo{cost function} constrains the demixing filters \RevFour{\eqref{eq:convolution_matrix}} to realize a linear convolution. \RevFour{This is expressed by the window matrices $\mathbf{U}$ in \eqref{eq:TRINICON_costFct_woKL} and in its frequency-domain representation \eqref{eq:TRINICON_costFct_DFT}.}

The limitations of the circular convolution model has been extensively studied in literature, e.g., \cite{kellermann_wideband_2003}, and especially for \ac{FD-ICA} in, e.g., \cite{araki_fundamental_2003,makino_blind_2005,sawada_convolutive_2004,pedersen_survey_2007}. To approximately represent linear convolution, i.e., to limit circular convolution effects, long \RevFour{\ac{STFT}} windows are desirable. However, this reduces the number of available data samples per frequency bin to estimate statistics needed for the \ac{BSS} algorithm \RevTwo{or requires a large degree of overlap of data blocks with accordingly increased computational complexity}. 
%Another drawback of very long \ac{DFT} windows is that the short-time stationarity assumption for speech signals may be violated, which implies the \RevTwo{inadequacy of the implicit} ergodicity assumption, which eventually renders the estimation of the statistics \RevTwo{questionable}. 
\RevTwo{As} \ac{FD-ICA} and \ac{IVA} share the same convolution model, \RevTwo{these} arguments \RevTwo{apply} to \ac{IVA} as well. 

Hence, we can conclude that \ac{FD-ICA} and \ac{IVA} \RevTwo{necessarily need to compromise regarding} the approximation of the linear convolution and the capability to estimate desired signal statistics. On the \RevTwo{other hand, no such compromise is necessary with \ac{TRINICON}.}
\begin{figure}
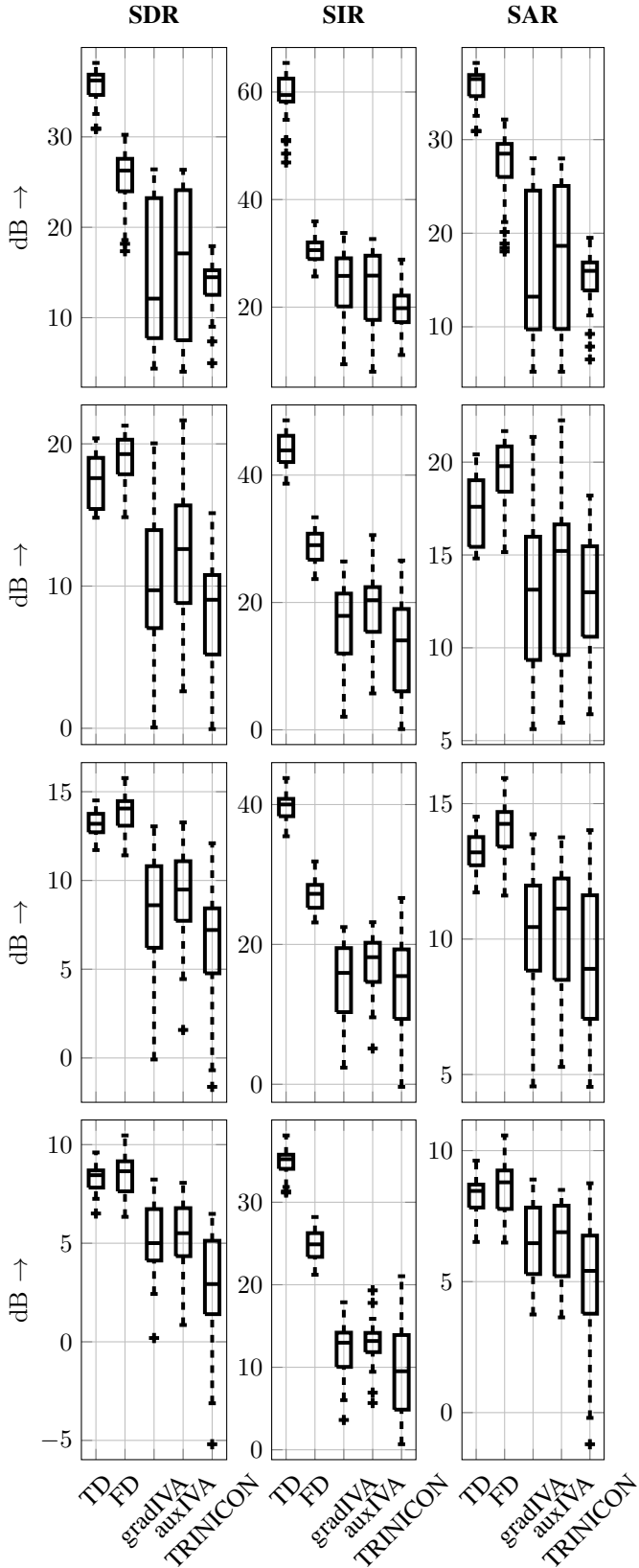

	\setlength\fwidth{0.4\textwidth}
	\pgfplotsset{every axis plot/.style={line width=1.5pt}} 
	\input{figures/HAR}\\ \input{figures/MMRcc}\\ \input{figures/MMRco}\\ \input{figures/R517}
	\caption{Results in terms of output \ac{SDR}, \ac{SIR} and \ac{SAR} of the discussed methods. The first row of plots corresponds to the low-reverberant chamber ($T_{60} = 0.05\,\mathrm{s}$), the second and third row to the two meeting rooms ($T_{60} = 0.2\,\mathrm{s}$ and $T_{60} = 0.4\,\mathrm{s}$) and the last row to the seminar room ($T_{60} = 0.9\,\mathrm{s}$).}
	\label{fig:results}
\end{figure}
%%%%%%%%%%%%%%%%%%%%%%%%%%%%%%%%%%%%%%%%%%%%%%%%%%%%%%%%%%%%%%%%%%%%%%%%%%%%
\subsection{Experiments}
\label{sec:experiments}
%%%%%%%%%%%%%%%%%%%%%%%%%%%%%%%%%%%%%%%%%%%%%%%%%%%%%%%%%%%%%%%%%%%%%%%%%%%%
\RevTwo{For \RevFour{an} experimental comparison of the methods \RevFour{which exemplarily illustrates the findings above}, we focus on \ac{IVA} and \ac{TRINICON} in a determined scenario of \RevThree{$P=2$} sources and two microphones \RevFour{and consider only the performance after convergence}. A comparison of the separation performance of \ac{FD-ICA} and \ac{IVA} has been presented in \cite{vincent_performance_2006}. The microphone signals are simulated by \RevFour{(linearly)} convolving speech signals randomly chosen from a set of four male and four female speakers of $10\,\mathrm{s}$ duration with measured \acp{RIR}. To obtain \RevThree{representative} results, in each experiment with a fixed acoustic situation, the two speech signals have been drawn $20$ times from the set of available clean speech signals. The \acp{RIR} were measured with a microphone pair of $4.2\,\mathrm{cm}$ spacing from four different rooms: a low-reverberant chamber ($T_{60} = 0.05\,\mathrm{s}$), two meeting rooms ($T_{60} = 0.2\,\mathrm{s}$ and $T_{60} = 0.4\,\mathrm{s}$) and a seminar room ($T_{60} = 0.9\,\mathrm{s}$). The sources have been placed at $1\,\mathrm{m}$ distance from the microphones and at $\pm 45^\circ$ from broadside direction.}

\RevTwo{To obtain a benchmark for ideal performance of a demixing system realized by circular and linear convolution, we identified Relative Transfer Functions (RTFs) between the two microphones based on the source images. On one hand, we estimated a relative impulse response as an \ac{FIR} filter realizing a linear convolution by solving a least squares problem in time domain. This approach is abbreviated as TD. On the other hand \RevFive{we identified} an \ac{RTF} separately in each frequency bin by solving a least squares problem, which represents a circular convolution and is named FD. To compare \ac{IVA} and \ac{TRINICON}, we chose natural gradient update schemes \cite{aichner_acoustic_2007,kim_blind_2007} and\RevFour{,} additionally for \ac{IVA}\RevFour{,} update rules derived from the \ac{MM} perspective called auxIVA \cite{ono_stable_2011}. For \ac{IVA}, a Laplacian source prior \cite{kim_blind_2007,ono_stable_2011} has been used. The exploitation of nongaussianity did not yield a significant benefit in the presented experiments in terms of the \RevFour{steady-state} performance of \ac{TRINICON} and has therefore been dropped, i.e., we used a \ac{SOS} realization of \ac{TRINICON} as described in Sec.~\ref{sec:trinicon_sosMethods}. \RevFour{However, as mentioned earlier, exploiting \ac{HOS} increases \RevFive{-- especially initial --} convergence speed (see \cite{buchner_blind_2003}).}}

\RevTwo{The filter length (corresponding to the \ac{DFT} length for the \ac{STFT}-based methods) of all approaches has been set to $2048$ at a sampling rate of $16\,\mathrm{kHz}$. For \ac{IVA} we chose a Hamming window with $50\%$ overlap for the \ac{STFT} transform. \RevFive{With $10\,\mathrm{s}$ signal duration and $f_s = 16\,\mathrm{kHz}$ sampling rate, this leads to $N=156$ frames for \ac{IVA} and $N=78$ frames for \ac{TRINICON}.} Furthermore, we used a step size of $0.1$ and $1000$ for gradient-based \ac{IVA} (gradIVA) and a step size of $0.005$ and $1000$ iterations for \ac{TRINICON}. For auxIVA, we used $100$ iterations. These parameters have been identified experimentally and showed good performance on average.
The performance of the algorithms has been measured in terms of \acf{SDR}, \acf{SIR} and \acf{SAR}~\cite{vincent_performance_2006}.}

\RevTwo{Fig.~\ref{fig:results} shows the results of the described experiments, where each row corresponds to one of the four considered acoustic enclosures. It can be observed that the oracle approaches TD and FD obtained as expected the best results, where the linear convolutive method TD outperformed FD in terms of \ac{SIR}. This is to be expected as the linear convolution model is physically more meaningful and follows the experimental setup. Furthermore, the clear superiority in terms of \ac{SIR} can be understood as the interference suppression was the objective we optimized for. The inferior performance in terms of \ac{SDR} and \ac{SAR} for large reverberation times \RevFive{can be attributed} to the \RevFive{fact that \acp{RTF} may require a longer time-domain representation than the impulse responses itself}. For the \ac{BSS} methods based on the natural gradient, \ac{TRINICON} and gradIVA obtained for small reverberation times similar performance. At larger reverberation times, \ac{TRINICON} has been outperformed by gradIVA. \RevFive{This can be explained by the fact that the exploited signal properties are less affected by reverberation in frequency domain than in time domain (see \cite{brendel_diss} for a detailed analysis).} \RevFour{By exploiting} the \ac{MM} principle\RevFour{,} auxIVA obtained better results than the gradient-based algorithms in all cases.}

\RevTwo{Although \ac{TRINICON} represents a more general cost function and uses a linear convolution model, which has been experimentally shown to be superior over the circular one, \RevFour{the performance after convergence is} slightly worse than gradIVA and significantly \RevFour{inferior to} auxIVA. Therefore, we conclude that future work should investigate more powerful ways for the optimization of the \ac{TRINICON} cost function that allow to exploit the generality of the cost function and the benefit of the linear convolution model.}
%%%%%%%%%%%%%%%%%%%%%%%%%%%%%%%%%%%%%%%%%%%%%%%%%%%%%%%%%%%%%%%%%%%%%%%%%%%%
\section{Conclusion}
\label{sec:conclusion}
%%%%%%%%%%%%%%%%%%%%%%%%%%%%%%%%%%%%%%%%%%%%%%%%%%%%%%%%%%%%%%%%%%%%%%%%%%%%
In this contribution, we provided an in-depth and rigorous analysis of the relation between the cost functions of \RevThree{various} convolutive \ac{BSS} approaches based on \ac{ICA}. The similarity of the derivation of the considered cost functions has been demonstrated by explicitly performing corresponding calculations. By deriving a common representation for all cost functions, we showed that the \ac{TRINICON} cost function includes the \ac{IVA} cost function, which again includes the \ac{FD-ICA} cost function as special cases. The main differences between the cost function we identified are the linear instead of a circular convolution model and the \RevTwo{exploitation of nonstationarity} in \ac{TRINICON}. Furthermore, the relation between the \ac{TRINICON} cost function and \ac{SOS}-based methods has been shown. These identified differences are discussed and related to results from the literature. In this sense, we provided a common roof for all of these algorithms allowing the identification of similarities and differences in the underlying models. \RevTwo{Finally, experiments using measured \acp{RIR} compared the performance of \ac{IVA} and \ac{TRINICON} and benchmarked it with oracle demixing matrices based on a linear and a circular convolution model, respectively.}
\bibliographystyle{IEEEtran}
\bibliography{literature_new_machine}

\end{document}